\documentclass[preprint,twocolumn,3p]{elsarticle}

\usepackage{amsmath,amssymb,amsfonts}
\usepackage{algorithmic}
\usepackage{graphicx}
\usepackage{textcomp}
\usepackage{xcolor}
\usepackage[ruled]{algorithm2e} 
\usepackage[nolist]{acronym}
\usepackage{soul}
\usepackage{marginnote}
\usepackage{lipsum}
\usepackage[inline]{enumitem}
\usepackage{hyperref}
\usepackage{array}
\usepackage{multirow}
\usepackage{tabularx}

\newcommand{\myssec}[2]{%
  \subsection{#1}%
  \label{sec:#2}%
}
\newcommand{\rsec}[1]{%
  Sec.~\ref{sec:#1}%
}

\newcommand{\added}[1]{%
  {\color{blue}#1%
  }%
}

\newcommand{\removed}[1]{%
  {\color{red}\st{#1}%
  }%
}
\renewcommand{\added}[1]{#1}
\renewcommand{\removed}[1]{}

\newcommand{\myfigeps}[3][width=3in]{%
\begin{figure}[htbp]%
\centering%
\includegraphics[#1]{figures/#2}%
\caption{#3}%
\label{fig:#2}%
\end{figure}%
}
\newcommand{\myfigfulleps}[3][width=\textwidth]{%
\begin{figure*}[t]%
\centering%
\includegraphics[#1]{figures/#2}%
\caption{#3}%
\label{fig:#2}%
\end{figure*}%
}

\newcommand{\rfig}[1]{Fig.~\ref{fig:#1}}

\newcommand{\rtab}[1]{Table~\ref{tab:#1}}
\newcommand{\req}[1]{Eq.~(\ref{eq:#1})}

\newenvironment{mylist}%
{%
\begin{enumerate*}[label=(\roman*)]%
}%
{%
\end{enumerate*}%
}

\newcommand{\internetref}[1]{%
\footnote{\url{#1}, accessed on Sep.~30, 2019.}%
}

\begin{document}

\title{%
Uncoordinated Access to Serverless Computing in MEC Systems for IoT

}

\author[1]{Claudio Cicconetti\corref{cor1}}%
\ead{c.cicconetti@iit.cnr.it}

\author[1]{Marco Conti}
\ead{m.conti@iit.cnr.it}

\author[1]{Andrea Passarella}
\ead{a.passarella@iit.cnr.it}

\cortext[cor1]{Corresponding author}
\address[1]{IIT, National Research Council, Pisa, Italy}

\begin{abstract}
  Edge computing is a promising solution to enable low-latency \ac{IoT}
applications, by shifting computation from remote data centers to
local devices, less powerful but closer to the end user devices.
However, this creates the challenge on how to best assign clients
to edge nodes offering compute capabilities.
So far, two antithetical architectures are proposed: centralized
resource orchestration or distributed overlay.
In this work we explore a third way, called uncoordinated access,
which consists in letting every device exploring multiple opportunities,
to opportunistically embrace the heterogeneity of network and load
conditions towards diverse edge nodes.
In particular, our contribution is intended for emerging serverless
\ac{IoT} applications, which do not have a state on the edge nodes
executing tasks.
We model the proposed system as a set of M/M/1 queues and show that
it achieves a smaller \removed{jitter} \added{delay} than single edge node allocation.
Furthermore, we compare uncoordinated access with state-of-the-art
centralized and distributed alternatives in testbed experiments
under more realistic conditions.
Based on the results, our proposed approach, which requires a tiny
fraction of the complexity of the alternatives in both the device
and network components, is very effective in using the network
resources, while incurring only a small penalty in terms of increased
compute load and high percentiles of delay.

\end{abstract}


\maketitle

  \section{Introduction}%
  \label{sec:introduction}%
  Nowadays \textit{edge computing} is a trending architecture where
applications on user devices are provided with computational
capabilities made available in the access networks.
Compared with traditional \ac{MCC}, edge systems enjoy lower latencies
and reduced Internet traffic.
These advantages make them desirable in several vertical market
segments, including mobile \ac{AR}/\ac{VR}~\cite{IGR2017}, connected
car~\cite{5GAA2017}, and \ac{IoT}~\cite{Automation2019}

Meanwhile, a new paradigm, called \textit{serverless computing} or
\ac{FaaS}, is also revolutionizing \ac{IoT} frameworks~\cite{Nastic2017}.
In serverless computing, processing is offloaded from the user
device by means of tasks similar to remote function calls, often
called \textit{lambda functions}, which are processed by remote
executors in a stateless manner~\cite{Varghese2018}.
Serverless computing was born as a cloud computing technology to
allow an easier up/down scaling of the executors in a data center
since there is no server-side state to be handled.
However, this paradigm fits very well many \ac{IoT} applications
that natively consist of event-driven or periodic execution of
processing jobs on data acquired in real-time for monitoring
purposes~\cite{Hussain2019}.

\myfigeps[scale=0.4]{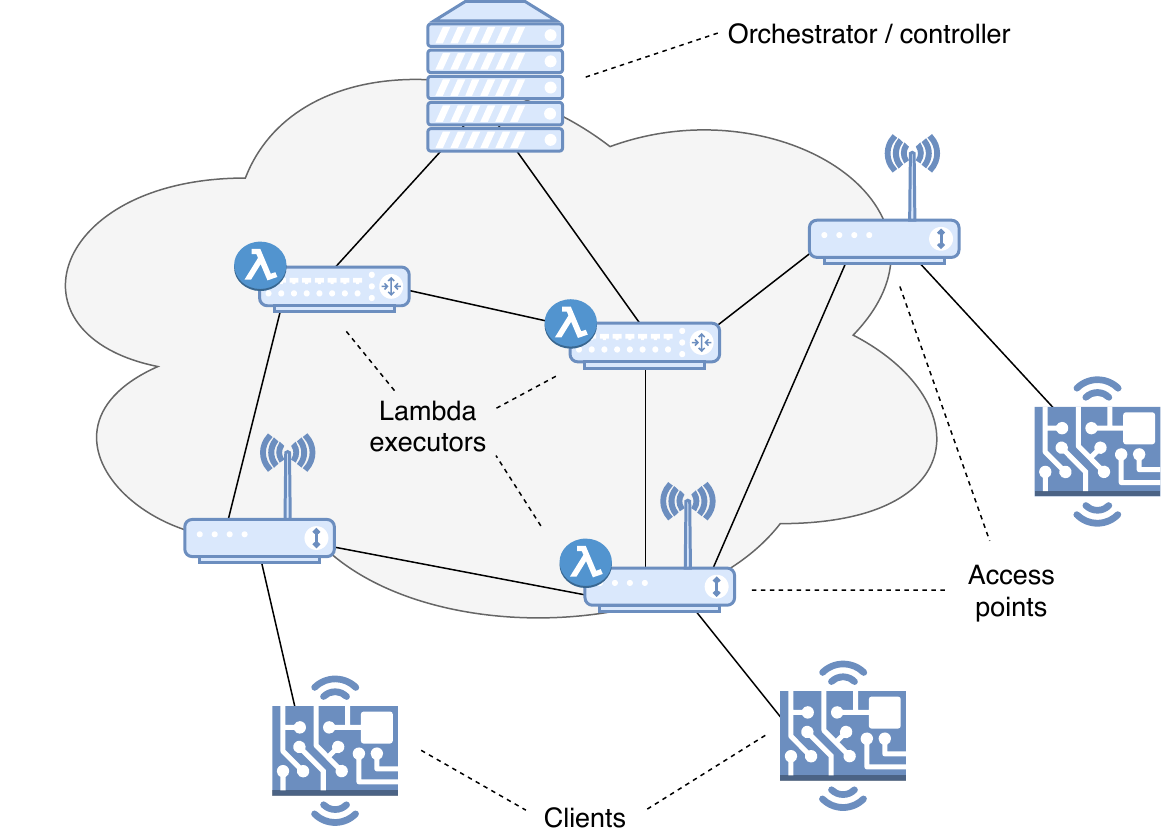}{Target scenario.}

In this paper we consider a system for \ac{IoT} applications that
combines the advantages of edge systems and serverless computing.
Our target scenario is illustrated by means of the example
in \rfig{target-scenario}, which shows an edge domain consisting
of:
\begin{mylist}
  \item access points, which provide the client devices with access
  to the edge network;
  \item lambda executors, co-located with network devices, which
  are equipped with spare/extra compute capabilities to respond to
  function execution requests from the clients;
  \item clients, which offload their computation by means of
  lambda requests towards the executors;
  \item a logically centralized entity, indicated as
  controller / orchestrator, which manages the lifecycle of the
  lambda images on the executors and dispatches the lambda
  requests from clients.
\end{mylist}
In the literature there are two alternative approaches for this
scenario, which will be analyzed in \rsec{contribution:arch}:
\textit{centralized}, where the association between clients and
executors is decided by the orchestrator, and \textit{distributed},
where the edge nodes cooperate to dispatch lambda functions from
clients to executors without a central coordination.
Both approaches require that a network-wide infrastructure is created
and maintained, which may incur a significant overhead and prove
inflexible to fast changing conditions, especially if the capabilities
of the lambda executors are limited, which is a use case of interest
for \ac{IoT} applications on low-power gateways in the network.

Therefore, to overcome the limitations of these existing approaches,
we propose to provide the clients with \textbf{uncoordinated access}:
a light orchestration assigns every client a pool of possible
lambda executors, but the final choice on where to send each and
every request is made by the client.
We discuss this solution in \rsec{contribution}, where we also
propose a practical decision mechanism based on probing of the
response time from different executors, which is simple enough to
be implemented even in \ac{IoT} devices with very limited computation
resources on board.
Furthermore, we show that our proposed solution is compatible \removed{with
minor changes} with the \ac{ETSI} \ac{MEC} standard~\cite{Taleb2017a},
which is attracting a growing interest from the edge computing
industry, especially in the mobile telco domain.
We provide the reader with a tutorial introduction to the standard
in \rsec{soa:etsi-mec}.

Furthermore, in \rsec{model} we model the proposed system under
simplifying assumptions, which allows us to perform a numerical
analysis of our proposed solution in the same section.
Finally, in \rsec{eval} we validate the conclusions obtained via
analysis of experimental results obtained with a proof-of-concept
implementation.
Experiments are carried out in an emulated network, configured with
realistic topology and traffic conditions, also comparing our
uncoordinated access to centralized and distributed solutions from
the literature.
  \section{State of the art}%
  \label{sec:soa}%
  The goal of this section is two-fold.
On the one hand, in \rsec{soa:distributed} we provide the reader
with an overview of the recent studies in the scientific literature
that are most relevant \added{to this work} \removed{to the realization of serverless computing
in an edge system for IoT use cases.}
\added{%
We note that, to the best of our knowledge, there are no works in
the literature that address specifically the topic of serverless
computing in edge systems for IoT; thus, we survey a selection of
works that, in our opinion, provide an adequate technical background
or ancillary solutions \textit{in preparation of addressing} the
challenges ahead.
}
On the other hand, in \rsec{soa:etsi-mec} we introduce the \ac{ETSI}
\ac{MEC} standard, which is relevant in its possible role as a
leading technology to deploy interoperable edge systems and
applications.

\myssec{Distributed computing in edge systems}{soa:distributed}

The amount of literature that could be ascribed to \ac{IoT} is
titanic.
After the outburst of works on \acp{WSN} more than 20 years ago,
our research community has produced architectures, protocols, and
algorithms for all possible requirements, some of which have made
it into standards and products in the market.
However, conclusive solutions have yet to be found regarding some
crucial aspects that still hinder the full potential of mass
applications to be unlocked, which is expected to pass through edge systems
thanks to the advantages they offer compared to both on-device
execution and pure cloud offloading, as already discussed in the
introduction.
These aspects include scalable and sustainable strategies for the
operation and continuous optimization of resources under realistic
assumptions, for which we illustrate the recent state of the art
in the following.
The interested reader may find further sources of inspiration in
the recent survey papers~\cite{Abbas2017, Porambage2018}.

The authors in~\cite{Hall2019} focus on the server-side implementation
challenge if having multiple services on an edge node with compute
capabilities, requiring isolation and low overhead, especially lower
than that imposed by full-fledged virtualization systems intended
for high-end servers.
To this aim, they propose to use WebAssembly, which is a binary
instruction format intended for applications to be executed with
native speed within web browser, but could equally be used as a
form of extremely down-scaled virtualization for \ac{IoT} services,
with similar goals as
Unikernels~\cite{Madhavapeddy:2013:URV:2557963.2566628}.
An even more further looking solution is proposed
in~\cite{Tasiopoulos2019}, where the micro-services are assumed to
be dynamically distributed and executed based on peer-to-peer
monetary incentives, which is a direction already pursued in the
market, e.g., in the Golem network
project\internetref{https://golem.network/}, though in the context
of \ac{HPC}.
In any case, our work builds on top of any such approach that allows
edge nodes to provide \ac{FaaS} micro-services that respond to
requests from clients: we aim at optimizing their access in the
short-term (seconds to minutes), while long-term optimizations will
be done regularly as part of the system's house-keeping activities.

Fault tolerance is the subject of some works,
including~\cite{Schaefer2018}, where distributed computing is
realized by means of so-called \textit{tasklets}.
The underlying assumption there is that executors are inherently
error-prone, because they are hosted on devices owned by (cooperative)
end users.
While this assumption may not apply to typical \ac{IoT} scenarios,
where failure of an edge node is expected to be a sporadic event,
we note that our proposed uncoordinated serverless access goes
exactly into the same direction, since it embeds reliability by
using a pool of executors rather than a single one.
Another problem that has attracted some interest recently is deciding
on the user device whether a given task should be offloaded to
edge/fog/cloud nodes or it would be better executed on the local
compute resources.
As in~\cite{Sthapit2018}, this problem usually creates trade-offs
between execution time and energy consumption, which fits very well
the use cases where the clients are smart phones.
On the other hand, a basic assumption of this work is that the
\ac{IoT} user devices have very limited computation capabilities
for taking sophisticated decisions, and even more so for executing
tasks by themselves.
However, we inspire from that work for the definition of the
mathematical system model in \rsec{model}.
An alternative solution has been also proposed in~\cite{Liu2019}
for the same problem, where a near-optimal decision algorithm based
on Q-learning is proposed.

Finally, we cite here the two alternative architectures mentioned
briefly in the introduction, which will be studied in more details
in \rsec{contribution:arch}: centralized vs.\ distributed.
A centralized solution, where a single logical entity implements
load-balancing on the client requests towards a pool of executors,
is the standard approach in all cloud-based serverless environments,
which have been evaluated, e.g., in~\cite{Mohanty2018} (open source)
and~\cite{Lynn2017} (commercial).
On the other hand, in our previous work~\cite{Cicconetti2018}, we
have proposed to distribute load balancing on the edge nodes
themselves to overcome the limitations of a centralized structure
in an irregular edge network.
However, our previous solution was not intended for \ac{IoT}
scenarios, where both edge nodes and clients may have limited
capabilities.
In \rsec{eval} we compare in a large-scale scenario the access
scheme proposed in this paper to both such approaches.

\myssec{ETSI MEC}{soa:etsi-mec}

The \ac{MEC} industry study group was founded in \ac{ETSI} in
2014\footnote{The original name of the committee was \textit{Mobile}
Edge Computing, later changed to \textit{Multi-access} Edge Computing
in accordance with the paradigm shift towards a technology-agnostic
set of specifications, intended for not only mobile wireless
networks.} to create an open environment for the deployment of
interoperable applications from all the actors in the edge ecosystem:
vendors, service providers, third parties.
\removed{%
Despite the increasing interest in the industrial community (at the
time of writing the committee counts almost 100 members), the
scientific community has not yet produced a substantial amount of
literature around this standard, for which two major releases have
been published already.
This is also one of the reasons why we linger quite extensively on
the topic, even though it is only collateral to the core contribution
of our work, which is general and does not rely specifically on
the architecture or mechanisms provided by the ETSI MEC.
}
As a matter of fact, most scientific works focus on specific aspects
of the standard.
In~\cite{Sabella2018} the authors show how a real-time video streaming
application may benefit form radio-level information provided through
the \ac{ETSI} \ac{MEC} interfaces, thus allowing a service provider
to improve the \ac{QoE} of its users through the use of open interfaces
(today the only option would be to sign a contract with every mobile
network operator, then use different proprietary interfaces to
gather information from each).
In the core network of a mobile operator \ac{SDN} and \ac{NFV} are
the state-of-the-art solutions for the deployment and operation of
services; in~\cite{Schiller2018} the authors explore their relationship
with \ac{ETSI} \ac{MEC}, which is a key aspect in a production
network.
The same problem is also addressed in~\cite{Huang2018}, with a
specific focus on redirecting traffic from a mobile user to its
edge node in a transparent manner during roaming.
However, it is yet to be understood whether \ac{SDN}/\ac{NFV} are
also relevant for \ac{IoT} systems where the devices for both
connectivity and computation are expected to be more heterogeneous
and with limited capabilities.

\myfigeps[width=2.5in]{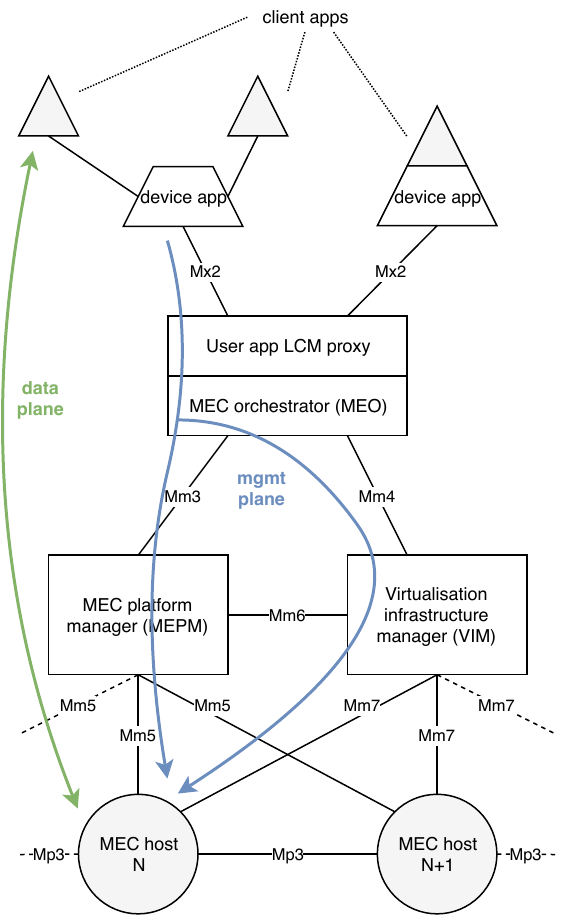}{Simplified blueprint of the ETSI MEC reference architecture.}

In the following we provide a short introduction to the standard,
with a focus on the aspects that are more relevant to this work.
\removed{%
The reader who is already familiar with the specifications may
safely proceed to the next section.
}
In \rfig{etsi-mec} we show the reference architecture of \ac{ETSI}
\ac{MEC}, as of release \removed{2.1.1} \added{2.1.2 (in draft at
the time of writing)}, with some simplifications regarding interfaces
and components that are not relevant to the discussion in this
paper.
The interested reader is referred to~\cite{Giust2018}, which provides
a complete overview of the standard, or directly to the \ac{ETSI}
\ac{MEC} specifications, which are all available to the public from
the group website\internetref{https://www.etsi.org/committee/1425-mec}.

In the upper part of the figure we show the client and device apps.
The \textit{client app} is \ac{ETSI} \ac{MEC} agnostic and it
interacts on the data plane with a \textit{user app} that physically
resides on a \ac{MEC} host.
The interface between the client and user app is application-dependent;
generally, serverless computing is carried out by means of
micro-services, hence the client only needs to know the \ac{URL}
or end-point of the lambda executor to offload computation tasks
to it.

On the other hand, the \textit{device app} is an \ac{ETSI} \ac{MEC}
aware component of the application running on the user device, which
interacts with the \ac{MEC} platform on the management plane through
the User Application \ac{LCM} Proxy using the \texttt{Mx2} \ac{API}.
The latter, as all the other \ac{ETSI} \ac{MEC} interfaces, is a
vendor-neutral RESTful interface, whose commands and data structures
are specified to facilitate interoperability
between application and platform software, intended to
be developed by different players in the ecosystem.
\removed{%
Using the \texttt{Mx2} interface a device app may:
(i) list the applications offered by the edge system;
(ii) create a context for the execution of a given application,
as needed by the client app;
(iii) delete an existing context when not needed anymore by the
client app.
The LCM proxy is a mere intermediate between the device apps
and the MEO, which is in charge of managing and optimizing the
use of resources across the whole MEC domain.
It does so by interacting with the MEPM and VIM, which
provide the MEO with a consistent management interface towards
the MEC hosts.
As can be seen in the bottom part of the figure, in addition
to the interfaces with the MEPM and VIM, the MEC
hosts also have an API for the communication with peers, e.g.,
to realize distributed resource optimization protocols, but this
opportunity has not been elaborated in detail so far in the standard.
}
The workflow expected from an application wishing to
use an \ac{ETSI} \ac{MEC} service is the following:
\begin{enumerate}[leftmargin=*]
  \item the client app invokes computation offloading via a proprietary interface
  on its device app; note that the client
  and device app may reside in different devices, e.g., in an \ac{IoT}
  system the client app may be in the smart object and the device app
  on a concentrator or gateway;
  \item the device app checks the availability of the application
  requested and initiates the creation of an application context;
  \item the \ac{MEO} checks the availability of resources and, if
  the new application is accepted, it allocates the necessary
  resources via the \ac{MEPM} and \ac{VIM} using the \texttt{Mm3}
  and \texttt{Mm4} interfaces, respectively; the algorithms and
  criteria used by the \ac{MEO} are voluntarily left open by
  the standard to foster market differentiation;
  \item these requests, in turn, reflects on the \ac{MEC} hosts via the
  \texttt{Mm5} and \texttt{Mm7} interfaces, respectively for
  computation and connectivity resources;
  \item once this flow on the management plane is completed, the
  device app is notified on the \texttt{Mx2} interfaces and the
  client and user app can start data plane interactions.
\end{enumerate}
At any time the \ac{MEO} can change the \ac{MEC} host serving the
client app for optimization reasons by means of a push notification
to the device app, e.g., if the mobile device roams to another area
of the wireless network or if the computation/network conditions
change due to other applications.
With non-serverless applications, this also incurs a state migrations,
which in general is a complex and costly operation.
In \rsec{contribution:etsi-mec-changes} we will describe how 
\added{%
to implement serverless uncoordinated access in ETSI MEC.
}
\removed{%
ETSI MEC interfaces have to be adapted to suit the
requirements of our uncoordinated access solution proposed below.
}
  \section{Uncoordinated Serverless Access}%
  \label{sec:contribution}%
  In this section we describe our proposed architecture for uncoordinated
access of \ac{IoT} clients to serverless micro-services in an edge
system.
\added{%
We start by defining the key requirements of \ac{IoT} architectures,
in general, in \rsec{contribution:reqs}, based on which we propose
our so-called \textit{uncoordinated access} in \rsec{contribution:arch}.
We then illustrate in \rsec{contribution:algo} a simple stateless
algorithm that can be used by the clients in this architecture,
intended as a baseline for constrained devices.
Depending on the availability of extra computational resources on
the client devices, one can think of more sophisticated solutions,
which are the subject of our current investigations.
}
We believe that our contribution is general enough to be suitable
to several edge technologies and target deployments; to confirm our
statement, in \rsec{contribution:etsi-mec-changes} we show how to
realize the framework with the \ac{ETSI} \ac{MEC}.

\myssec{\added{Requirements}}{contribution:reqs}

Before delving into the illustration of our contribution, we elaborate
a moment on the following four \textbf{fundamental requirements}
that \textit{any} architecture should meet to be an effective
solution in our context:
\begin{enumerate}[leftmargin=*,label=\Alph*.]
  \item \textit{It should be easy to implement on the user side:}
  \ac{IoT} devices often have very limited CPU/memory capabilities.
  \item \textit{It should be lean on edge compute resources:} in
  an \ac{IoT} scenario the network infrastructure usually consists
  of WiFi \acp{AP} or other \ac{SoC} / low-power devices, which are
  equipped with specialized hardware, e.g., FPGAs or GPUs, that
  makes them suitable as servers for specific applications, but
  whose processing capabilities available for control / management
  activities is limited.
  \item \textit{It should adapt well to fast changing conditions:}
  in many use cases of practical interest the user devices are
  mobile and the application patterns are not known \textit{a
  priori}, thus it is not possible to optimize once and for all the
  allocation of clients to edge nodes.
  \item \textit{It should be lean on backhaul resources:} the
  connectivity of the edge nodes, both between themselves and
  with core network components, called \textit{backhaul} in telco
  terminology, may be scarce and heterogeneous.
  \item \textit{It should be cheap to maintain:} due to the sheer
  numbers of devices expected to be connected for future \ac{IoT}
  applications, we argue that any sustainable business model must
  severely limit the expenses for operating and monitoring a deployed
  infrastructure, which in most cases will grow over time and remain
  in place for much longer than, e.g., mobile wireless access
  infrastructures, which have to catch up every few years with
  constantly advancing technologies.
\end{enumerate}

\myssec{Proposed architecture}{contribution:arch}

\myfigeps[width=0.45\textwidth]{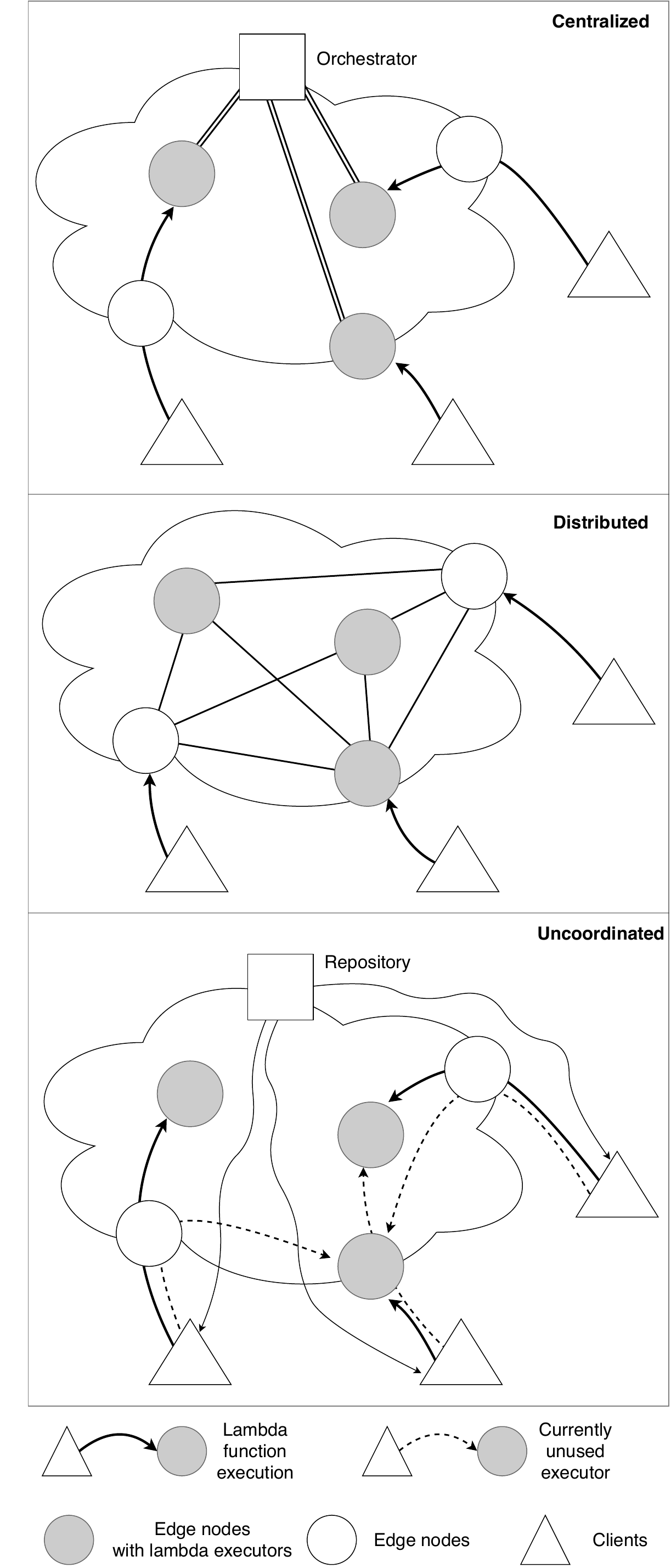}{%
Comparison of the proposed architecture for uncoordinated access
to serverless functions (bottom) with traditional centralized (top)
and distributed (middle) approaches.
%
%
%
%
}

Let us consider first \textbf{centralized solutions}, which are the
baseline approach in cloud-based serverless solutions, such as
Knative\internetref{https://knative.dev} or Apache
OpenWhisk\internetref{https://openwhisk.apache.org}, and telco-native
architectures, including the \ac{ETSI} \ac{MEC} as illustrated in
\rsec{soa:etsi-mec}.
With this paradigm, illustrated in the top part of \rfig{architectures},
the applications on the user devices merely obey to a logically
centralized orchestrator, which instructs them to which end-point
or \ac{URL} to address their lambda requests.
Since the entire decision making is done by the orchestrator alone,
requirements A and B above are automatically covered.
Also, since the orchestrator has a system view, we can expect that
it can follow very well the changing conditions, possibly even
anticipating such changes if prediction algorithms are used,
which meets requirement C.
For the same reason, requirement E is also addressed: the orchestrator
is the only complex component of the system that needs be monitored
and maintained.
However, centralized solutions fall short in covering requirement D:\@
making appropriate decisions require that the orchestrator is updated
by all edge nodes on the real-time status of its resources.
This may be
reasonable in cloud-based solutions, where all the executors are
powerful and well connected and in close proximity to one another
in a data center, but it certainly poses limits to the growth of
the edge system as the backhaul gradually becomes a choke point.

For this reason, in the literature some \textbf{distributed
solutions} have been proposed (see \rsec{soa:distributed}).
As illustrated in the middle part of \rfig{architectures}, the basic
concept of these approaches is that an overlay exists between the
user devices and the executors, made by edge nodes that take local
decision in a distributed manner to optimize the execution of
services, which greatly reduces the internal traffic thus
meeting requirement D.
The user devices remain unaware of the underlying complexity, hence
requirement A is met, as well.
Adaptation to changing conditions (requirement C) is addressed, as
long as the distributed system can reach near-optimum working point
despite the decision makers have a limited view of the system.
However, distributed solutions cannot address adequately requirements
B and E.
\added{%
On the one hand, taking informed decisions in a fully distributed
manner requires that the edge nodes coordinate among themselves
and dedicate part of their computational capabilities to the
process of maintaining a synchronized state for this purpose;
therefore requirement~B may be difficult to achieve, especially
with a high number of edge nodes in the system.
On the other hand, the edge system operator would have to maintain
a potentially large number of active components in the system,
including monitoring, supervision, and software upgrades phases,
also addressing heterogeneous hardware and software characteristics,
which makes it challenging to achieve requirement~E.
}
\removed{%
On the one hand, they require the edge nodes to have sufficient
capabilities in terms of CPU and memory to store and analyze data to
take informed decisions in a distributed manner (requirement B).
On other other hand, system operation requires maintaining several
components rather than a single one, with significantly increased
costs (requirement E).
}

In this work we propose a solution, called here \textbf{uncoordinated
access}, that overcomes the respective limitations of centralized
/ distributed approaches and addresses all the requirements.
We start with an observation: slow-changing conditions in the system
are easily detectable by a centralized entity, i.e., the orchestrator,
with a low overhead since this merely requires aggregate measures
from the executors and it is not a real-time task.
Thus, let us assume that the lifecycle of the micro-service images
on the executors is somehow optimized so as to follow \textit{macroscopic}
slow trends in the system.
The real challenge is following the \textit{microscopic} fast changes: if,
for instance, an executor is installed in a \ac{SoC} device, such as
a Raspberry Pi, then very few concurrent executions of a lambda
function can easily overload the executor, thus increasing the
response times of clients and possibly degrading the application.
Following these variations in a system with no reservation of
resources nor \textit{a priori} knowledge on the arrival of lambda
execution jobs is extremely challenging, and in fact leads to their
respective key shortcomings of the centralized and distribution
solutions discussed above.
On the other hand, rather than complicating the system to \textit{beat}
this variability, we propose to \textit{embrace} it: we propose
that the orchestrator allocates a pool of end-points / \acp{URL}
to every client, which the latter can use to exploit opportunistically
to its own advantage taking internal decisions.
This is illustrated in the bottom part of \rfig{architectures},
where solid lines represent the current choice of destination of
clients and dashed lines are the (currently) unused
alternatives they have been informed about by the central entity;
the latter is called here \textit{repository} to stress that it merely
communicates a pool of executors to every client without running a
real-time optimization process, as in a centralized architecture.

\added{%
We call this solution \textit{uncoordinated} because the clients
do not interact with other components to take decisions on a per
request basis.
In an environment that evolves with fast dynamics, relying on the
statistical multiplexing of uncoordinated agents taking myopic
``good enough'' decisions may result beneficial compared to a system
trying to achieve ``optimal'' goals, which however fails because
it is either fed outdated information or it consumes too many resources
(computation, traffic) in the process.
On the other hand, system-wide optimization can be added on top of
the proposed solution, working at a much slower time scale (in the
order of minutes and above).
This can be done along at least the following two directions:
modifying the set of executors deployed on edge nodes (e.g., an
algorithm based on popularity of functions requested was proposed
in~\cite{Krol2017}); advertising different pools of executors to
the clients, based on long-term estimates of networking and computation
statistics, which can be seen as a service placement problem
(see~\cite{Pasteris2019}).
}

\removed{
In other words, we argue that relying on the statistical multiplexing
in a changing environment is more efficient than trying to hit an
ever moving target, since the aiming process itself would consume
a significant amount of resources in terms of computation, network
traffic, maintenance.
We call this solution \textit{uncoordinated}, not distributed,
because the clients never interact with one another to take decisions.
}
\added{%
It is straightforward to see that the proposed uncoordinated solution
meets all the requirements in \rsec{contribution:reqs}, with the
following two minor notes.
}
First, this design only makes sense if the clients can decide
which executor to use in a simple manner (requirement A), as we
explain below in \rsec{contribution:algo}.
Second, as for distributed solutions, we have to abandon the goal
of achieving a global optimum, since this would require either a
system-wide view or an extremely complex/expensive synchronization
across the clients; however, we argue that ``good'' performance
levels in a practical solution are way more preferable than reaching
optimum performance under unfeasible conditions.

\begin{table*}[htbp]%
\caption{%
Qualitative comparison of the proposed architecture (uncoordinated)
with classical centralized and distributed approaches from the
literature, in terms of meeting five fundamental requirements (see
\rsec{contribution:reqs}).}%
\vspace{-2em}%
\begin{center}%
{\small%
\begin{tabularx}{0.8\textwidth}{|X|c|c|c|}
\hline
Requirement & Centralized & Distributed & Uncoordinated \\
\hline
A. Easy to implement on the user side & $++$ & $++$ & $+$  \\
B. Lean on edge compute resources     & $++$ & $-$ & $++$ \\
C. Adaptation to changing conditions  & $++$ & $+$  & $+$  \\
D. Lean on backhaul resources         & $--$ & $+$  & $++$ \\
E. Cheap to maintain                  & $++$ & $--$ & $++$ \\
\hline
\end{tabularx}
}%
\label{tab:architectures}%
\end{center}%
\end{table*}%

The above discussion is summarized for the readers' convenience in
\rtab{architectures}.

\myssec{\added{Client algorithm}}{contribution:algo}

To complete our proposition we now describe how the clients select
over time the executor to be used from the pool of those available.
\added{%
The pool of executors must be communicated to clients by a component
with system-level view, indicated as a \textit{repository} in
\rfig{architectures}, which in a real system would interact with
the orchestrator in charge of managing the life cycle of executors
on edge nodes.
The algorithm that is used by (e.g.) the orchestrator to decide
which pool of executors has to be notified to which client may be
subject to optimization too, and is a research issue \textit{per
se}.
However, since this happens at a time scale greater than that of
interest for the scheduling of lambda functions, we consider this
specific issue out of the scope of this work, and subject to future
investigations.
In the performance evaluation in \rsec{eval}, we select the pool
of executors that minimize the number of network hops for a client
to reach them, also limiting the pool size to 2 or 3.
}

\myfigeps[scale=0.4]{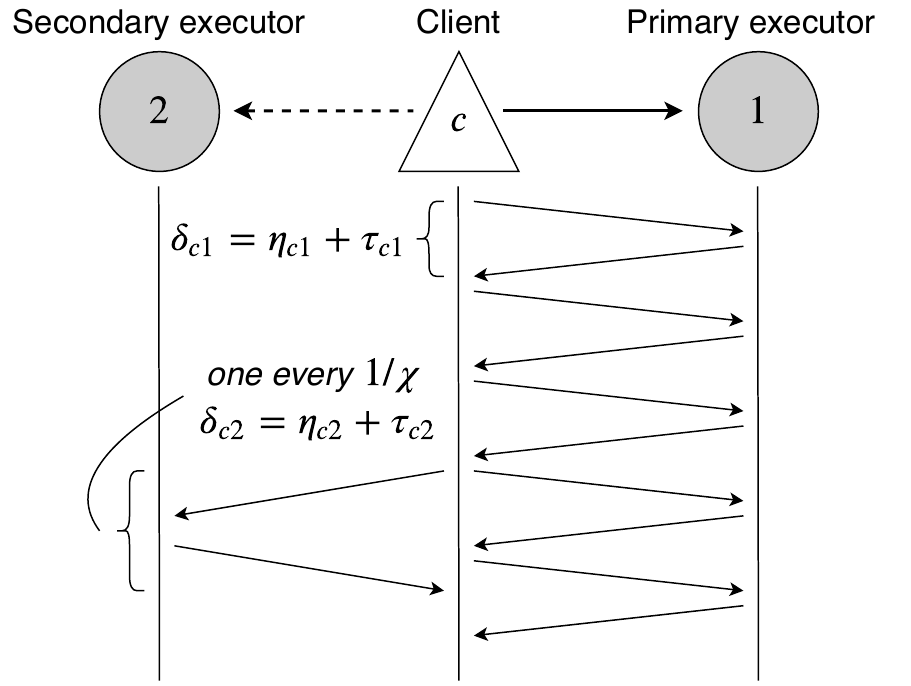}{%
Sequence diagram of the execution of lambda functions from a client
towards its primary executor and, once every $\chi$ requests on
average, to the secondary executor, as well, for probing purposes.
In the figure $\delta$ is the overall delay, consisting of processing
component $\eta$ and network component $\tau$; the full notation
is introduced in \rsec{model} and summarized in \rtab{notation}.}

To keep the client very simple, consistently with requirement A
above, we propose that it keeps one of the possible destinations
are the current one, then at every execution of the lambda function
it selects also another destination for probing with a given
probability $\chi$, which is a system parameter.
The lambda function is then issued both to the \textit{primary}
executor and to the \textit{secondary} one selected: after measuring
the relative latencies, the client may then promote the secondary
to primary.
\added{%
An example sequence diagram is shown in \rfig{seq-diagram}.
The primary executor (1) is depicted on the right of the client $c$,
which issues lambda function requests to 1 only with probability
$1-\chi$.
Sporadically, i.e., on average every $1/\chi$, one request
will be issued towards both the primary executor (1) and the secondary
executor (2), depicted on the left of $c$.
In the example, the overall response time from 2 is greater than
than from 1, hence the process continues at the client side as before.
}

In fact, the main reason for a network or service operator to use
edge computing is because the applications have latency constraints.
If this is not the case, then cloud computing is bound to be
a cheaper and easier alternative for economy of scale reasons.
We note that the latency of a lambda transaction consists of several
components, including the time to transmit the messages and the
responses, network queuing delays between any two hops, and the
lambda execution time plus any additional waiting due to the
application/OS scheduler.
However, from the point of view of the application on the user
device, such decomposition is irrelevant: what counts is only the
time between when the lambda function is issued and when a response
is received, which can be easily measured locally.

Finally, the reader may wonder at this point why the lambda function is
executed towards both destination instead of only the one under
probing: this is to make sure in the most simple manner that the
latency measurements are comparable.
In fact, not all tasks of the same lambda type may be the same,
e.g., the input may have a different size or contain data that are
more or less complex to process on the edge node side; furthermore,
environmental conditions may change from one lambda execution to
the next one, e.g., the access link of the client may suffer from
wireless impairments temporarily reducing the bit-rate.

%
%
\removed{%
This proposed mechanism is to be intended as a baseline, since it
is stateless on the client, which is merely required to be able to
fire two lambda functions together.
However, depending on the availability of extra computational
resources on the clients, we can think of more sophisticated
solutions, which are currently the subject of further investigations.
}

%
%
%
%
%

\myssec{ETSI MEC Implementation}{contribution:etsi-mec-changes}

We now describe how to implement the proposed uncoordinated serverless
scheme with \ac{ETSI} \ac{MEC}\removed{, along with the minor changes
required to its APIs.}
The reader is referred to \rsec{soa:etsi-mec} for a tutorial
introduction to this standard.

As mentioned already, with serverless computing the executors do
not hold a state for every active client application.
This property can be exploited by the \ac{MEO} to load the lambda
images, e.g., \ac{VM} or containers, on the \ac{MEC} hosts, according
to slow-changing estimations / predictions of their utilization,
which is outside the scope of this work.
This way, all the interfaces from \texttt{Mm3} to
\texttt{Mm7} (see \rfig{etsi-mec} above) are not used either for
execution of lambda transactions or for creation of new application
contexts from device apps.
The management plane, as such, is greatly simplified compared to
traditional (stateful) applications, and in fact the \ac{MEO} merely
acts a repository of the end-points of the available lambda images
on all the \ac{MEC} hosts, grouped per lambda function type.
Simplification also translates into a better scalability as the
rate of context creation increases, which is a very desirable
property in \ac{IoT} scenarios where we can expect that some
applications will have a short-lived duration.

As an application context creation from the device app is requested
on the \texttt{Mx2} interface, via the \ac{LCM} proxy, the \ac{MEO}
selects a number of executors and includes their end-points in the
response to the device app.
\removed{%
This requires a change to the response message itself, called
\texttt{AppContext} in the standard, since the latter only allows
a single address of the user application to be specified in
a field called \texttt{referenceURI}.
However, such modification is extremely simple to implement: for
instance, we could add another field \texttt{referenceURIlist} that
contains the pool of end-points, while copying one of them into the
existing \texttt{referenceURI} field to support backward-compatibility
with legacy device apps.
Similarly, we could add a new field \texttt{chi}, simply ignored
by legacy device apps, which instructs the device app on the amount
of overhead it is allowed to spend for probing purposes.
}

The algorithm by which the \ac{MEO} determines both the $\chi$ and
which executors are to be selected for every new context is beyond
the goal of this paper and part of our on-going research activities,
fostered by the model illustrated in \rsec{model} below as a building
block for the design of such an optimized algorithm in a production
system.
  \section{System Model and Analysis}%
  \label{sec:model}%
  In this section we present a mathematical model of the uncoordinated
serverless access system put forward in \rsec{contribution}, under
simplifying assumptions to make it tractable (\rsec{model:system}).
To facilitate the reader visualizing the model, we then study the
simple case of 2 clients served by 3 executors (\rsec{model:example}).
Finally, we provide numerical results to compare a static allocation
to our proposed solution and derive some system properties
(\rsec{model:analysis}).
The conclusions found will be validated against experimental results
in the next section \rsec{eval}.

\myssec{System model}{model:system}

Despite the simplicity of implementation in both the client and
the orchestrator, the proposed system is still too complex to
be formalized in mathematical terms in general conditions.
Therefore, we now make the following simplifying assumptions.
We assume to have a set of $\mathcal{C}$ clients, each issuing
lambda function requests of the same type towards a pool $\mathcal{E}$
of executors.
For simplicity, we assume that both the arrival rate of tasks at
every client and the serving rate at the executors are Poisson
distributed, with average $\lambda_i$ for client $i$ and $\mu_j$
for executor $j$.
We assume that the network delay $\tau_{ij}$ between any client $i$
and executor $j$ is constant and independent of the state of the
system.
Finally, we assume that the orchestrator provides every client with
exactly two possible choices, which we call primary and secondary
depending on which one is currently selected.

\begin{table*}[htbp]%
\caption{Notation used.}%
\vspace{-2em}%
\begin{center}%
{\small%
\begin{tabularx}{0.8\textwidth}{|r|r|X|}
  \hline
  \textbf{Symbol} & \textbf{Domain} & \textbf{Meaning} \\
  \hline
  $\mathcal{C}$ & $\left\{1, \ldots, C\right\}$ & Set of clients, $|\mathcal{C}| = C$ \\
  $\mathcal{E}$ & $\left\{1, \ldots, E\right\}$ & Set of executors, $|\mathcal{E}| = E$ \\
  $\mathcal{S}$ & $\left\{1, \ldots, 2^C\right\}$ & Set of states, $|\mathcal{S}| = 2^C$ \\
  $\mathbf{\tau}$ & $\mathbb{R}^{C \times E}$ & Network delays \\
  $\mathbf{x}$ & $\mathbb{R}^C$ & Task processor utilization \\
  $\mathbf{\lambda}$ & $\mathbb{R}^C$ & Task arrival rate \\
  $\mathbf{\mu}$ & $\mathbb{R}^E$ & Task dispatch rate \\
  $\mathbf{s}$ & ${\left\{1, \ldots, E\right\}}^{C \times S}$ & Primary executor per client per state \\
  $\mathbf{\bar{s}}$ & ${\left\{1, \ldots, E\right\}}^{C \times S}$ & Secondary executor per client per state \\
  $I_{ijk}$ & $\langle i, j, k \rangle \rightarrow \left\{0, 1 \right\}$ & Primary indicator function \\
  $\bar{I}_{ijk}$ & $\langle i, j, k \rangle \rightarrow \left\{0, 1 \right\}$ & Secondary indicator function \\
  $\mathbf{\delta}$ & $\mathbb{R}^{C \times S}$ & Average delay with primary executor per client per state \\
  $\mathbf{\bar{\delta}}$ & $\mathbb{R}^{C \times S}$ & Average delay with secondary executor per client per state \\
  $\chi$ & $\left(0, 1\right)$ & Probability that the secondary executor is probed \\
  $\mathcal{J}_k$ & $\subseteq \mathcal{S}$ & Set of states reachable by state $k$ \\
  $P$ & ${\left[0, 1\right]}^{S \times S}$ & Transition matrix of the associated DTMC \\
  $\mathbf{\pi}$ & ${\left[0, 1\right]}^{S}$ & Stationary distribution \\
  \hline
\end{tabularx}
}%
\label{tab:notation}%
\end{center}%
\end{table*}%

For consistency and better readability we adopt the following rules
in the notation:
\added{
the indices $i$ and $h$ always refer to clients, the index $j$
to executors, the index $k$ to states;
}
\removed{
the indices $i$ and $h$ / $j$ / $k$ and $z$ always refer to clients
/ executors / states, respectively;
}
vectors and matrices are indicated
in \textbf{bold} (e.g., $\mathbf{x}$) and their corresponding
elements use the same letter in regular font (e.g., $x_i$).
A summary of the notation used is reported in \rtab{notation}.

\myfigeps[scale=0.4]{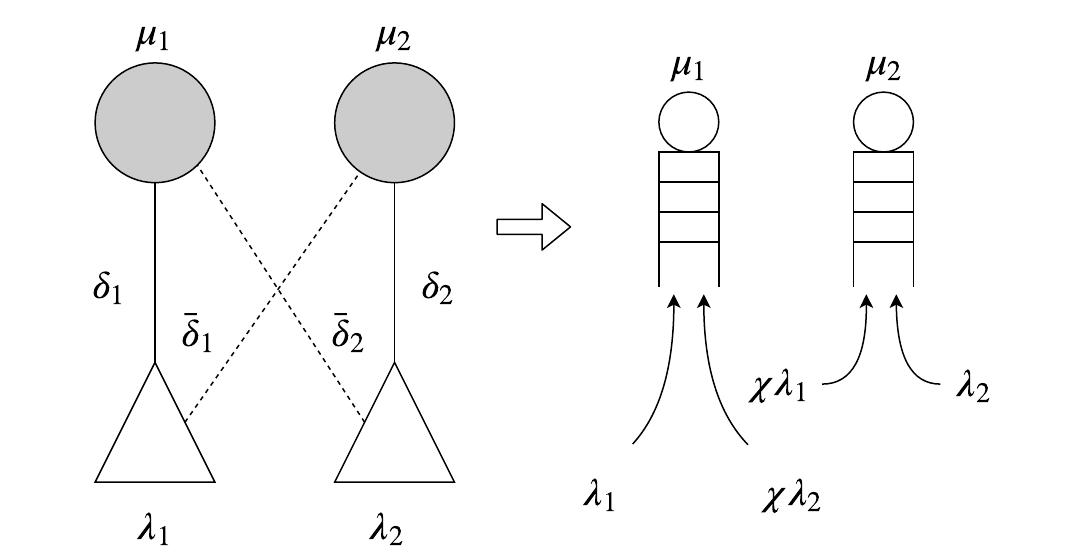}{%
System model, illustrated by means of an example with two clients
and two executors, with $\mathbf{\tau} = \mathbf{x} = \mathbf{0}$;
the equivalent M/M/1 model is on the right.}

Because of our assumptions above, we can consider the client and
executors as a set of Markov M/M/1 queue systems.
Without loss of generality we assume that the executors follow a
\ac{PS} policy, which we believe to best approximate how a real
edge node behaves under typical working conditions.
Thus, for each client $i$ we also define its processor share $x_i$.
The equivalent M/M/1 system is illustrated by means of the example
in \rfig{system-model} in the specific case where client 1 has
executor 1 as primary and 2 as secondary, while the opposite applies
to client 2.
We call the set of these conditions a \textit{state}, because
it captures one of the possible combinations in which our system
can be.
If, for instance, the executor 2 becomes primary for client 1,
then the system will be in a different state.
The set of the possible states $\mathcal{S}$ is given by the binary
enumerations of clients, because there are only two possible options
(primary vs.\ secondary), thus $|\mathcal{S}| = 2^C$.
In any specific state $k$, such as that depicted in \rfig{system-model},
we can identify the average delay $\delta_{ik}$ of any client $i$
towards its primary executor, as well as the same quantity towards
its secondary executor, called $\bar{\delta}_{ik}$.
It is important to note that for every executor the inbound tasks are both
those generated by the tasks that have it as the \textit{primary} destination and
those generated for probing reasons by the tasks that have it as the
\textit{secondary} destination.
The latter follow the same Poisson distribution, but are only a
fraction $\chi$ of the former.

According to basic queuing theory (e.g.,~\cite{Meyn1993}) the
average delay in an M/M/1 \ac{PS} system, including both queuing
and processing times, is:
\begin{equation}
  \label{eq:avg-delay-simple}
  \eta = \frac{x_i \mu}{\mu - \sum_h \lambda_h},
\end{equation}
where $\lambda_h$ is the arrival rate of any client $h$ served by
the executor, which only holds if $\mu > \sum_h \lambda_h$, i.e.,
if the system is \textit{stable}.
In our model we must take into account that, in any state $s$, an
executor serves only the clients having it as a primary or secondary
destination in that state.
To capture this property, we define the following two \textit{indicator
functions}.
First, $I_{ijk}$ is $1$ only if client $i$ has primary executor $j$
in state $k$, i.e., $s_{ik} = j$, otherwise it is $0$.
Likewise, $\bar{I}_{ijk}$ is $1$ only if $\bar{s}_{ij} = j$, otherwise
it is $0$.
We can now express the average delay of client $i$ towards its
primary executor when the system is in state $k$ as follows, based
on \req{avg-delay-simple}:
\begin{equation}
  \label{eq:avg-delay-primary}
  \delta_{ik} = \frac{%
    x_i \mu_{s_{ik}}}%
    {\mu_{s_{ik}} - \lambda_i -
      \displaystyle\sum_{h \in \mathcal{C}, h \ne i} \lambda_h \left[ 
        I_{hs_{ik}k} + \chi \bar{I}_{hs_{ik}k}\right]} +
  \tau_{is_{ik}},
\end{equation}
and, similarly, the average delay of client $i$ towards its secondary
executor when the system is in state $k$:
\begin{equation}
  \label{eq:avg-delay-secondary}
  \bar{\delta}_{ik} = \frac{%
    x_i \mu_{\bar{s}_{ik}}}%
    {\mu_{s_{ik}} - \chi \lambda_i -
      \displaystyle\sum_{h \in \mathcal{C}, h \ne i} \lambda_h \left[ 
        I_{h\bar{s}_{ik}k} + \chi \bar{I}_{h\bar{s}_{ik}k}\right]} +
  \tau_{i\bar{s}_{ik}},
\end{equation}
Both \req{avg-delay-primary} and \req{avg-delay-secondary} assume
that the queues are stable, i.e., that the respective denominator
is positive.
If this condition is not true, then the queue length grows over
time and the average delay tends to infinity in theory, while in
practice this condition will lead to much higher delays than usual.

Right up to this point we have shown how to build the two matrices
$\mathbf{\delta}$ and $\mathbf{\bar{\delta}}$, which give us the
average delays experimented by every client towards its two
possible destinations.
We now use this information to infer the average behavior of the
system at a steady state and, hence, derive the average delay
of every client.
Let us consider that the real system is dynamic: every client
randomly performs probing on the primary vs.\ secondary executor,
based on which it decides whether it should swap their role.
If we assume that every client takes decisions based on
the \textit{average} delay, as expressed in \req{avg-delay-primary}
and \req{avg-delay-secondary}, we see that the next state
\textit{for a given client $i$} is fully determined: if
$\delta_{ik} > \bar{\delta}_{ik}$, then $i$ will continue
using $s_{ik}$ as its primary executor; otherwise, the new
primary executor will become $\bar{s}_{ik}$.
However, this is an uncoordinated systems where all the clients take
their decisions individually, thus the transition from any state
$k_1$ to $k_2$ is determined by the random times when all the clients
take their swap decisions.
In other words, it is a stochastic process, which we can represent
by means of an associated \ac{DTMC}, where each state is exactly
one of the possible states $\mathcal{S}$ of our system, and the
transition matrix $P$ is built as follows.
For every state $k \in \mathcal{S}$ we consider all possible states
$\mathcal{J}_k$ that can be reached, where state $z \in \mathcal{J}_k$
iff the system can go from $k$ to $z$ with a combination of clients
$i$ changing their primary executor because of $\delta_{ik} >
\bar{\delta}_{ik}$:
\begin{equation}
  \label{eq:definition-jk}
  \mathcal{J}_k= \left\{ \forall z \in \mathcal{S}, z \ne k |
    \forall i : \left( \delta_{ik} \leq \bar{\delta}_{ik} \wedge s_{ik} = s_{iz} \right)
  \right\}
\end{equation}
To simplify notation, we assume that all the queues are stable, in
both the primary and the secondary executors\footnote{In the numeric
analysis in \rsec{model:analysis} below, we have taken into account
unstable queues as follows: $\delta_{ik} > \bar{\delta}_{ik}$ only
if $\delta_{ik}$ is finite (i.e., stable queue towards the primary
executor), in which case the condition is always true if
$\bar{\delta}_{ik}$ is infinite (i.e., unstable queue towards the
secondary executor).}.
If there is a state $k$ such that all the delays to the primary
executor are smaller than the delays to the secondary executor
(i.e., if $\exists k : \forall i, \delta_{ik} \leq \bar{\delta}_{ik}$),
then it is $\mathcal{J}_k = \emptyset$.
In this case $k$ is an \textit{absorbing state}; in our system this
means that every client sees the secondary destination as a worse
option compared to the primary destination, thus it does not swap
the two, and the system remains stable indefinitely.
In general, the cardinality of this set is given by:
\begin{equation}
  |\mathcal{J}_k| =
    2^{\left|\left\{\forall i, \delta_{ik} > \bar{\delta}_{ik}\right\}\right|} - 1
\end{equation}
It may also happen that state $k$ is \textit{unreachable}, i.e.,
$\forall z : z \notin \mathcal{J}_k$; an unreachable state will not
be reached unless the system starts from it.

Once all the $\mathcal{J}_k$ are determined for each $k \in
\mathcal{S}$, the transition probability $p_{kz}$ from state $k$
to state $z$ in the matrix $P$ is:
\begin{equation}
  \label{eq:transition-prob}
  \forall k,z \in \mathcal{S},
  p_{kz} = \begin{cases}
    0               & \text{if } z \notin J_k \\
    \frac{1}{|J_k|} & \text{if } z \in    J_k
  \end{cases}
\end{equation}
Without considering the systems with absorbing states, which are
of little practical interest to our analysis, and after removing
the unreachable states, we obtain a chain that is irreducible (i.e.,
it is possible to go from any state to any other), and whose states
are positive recurrent by construction.
Thus, the chain has a positive unique stationary distribution
$\mathbf{\pi}$, which gives the average probability in the long
term that the system will be in any given state.
Finally, we can use the latter to derive the average delays per
client as $\mathbf{\delta} \mathbf{\pi}$, i.e.:
\begin{equation}
  \label{eq:delay-final}
  E[\delta_i] = \displaystyle\sum_{k=1}^{S} \delta_{ik} \pi_k
\end{equation}
To better visualize the process and system variables we report
in the following a simple numeric example.

\myssec{Example}{model:example}

\myfigeps[width=2.5in]{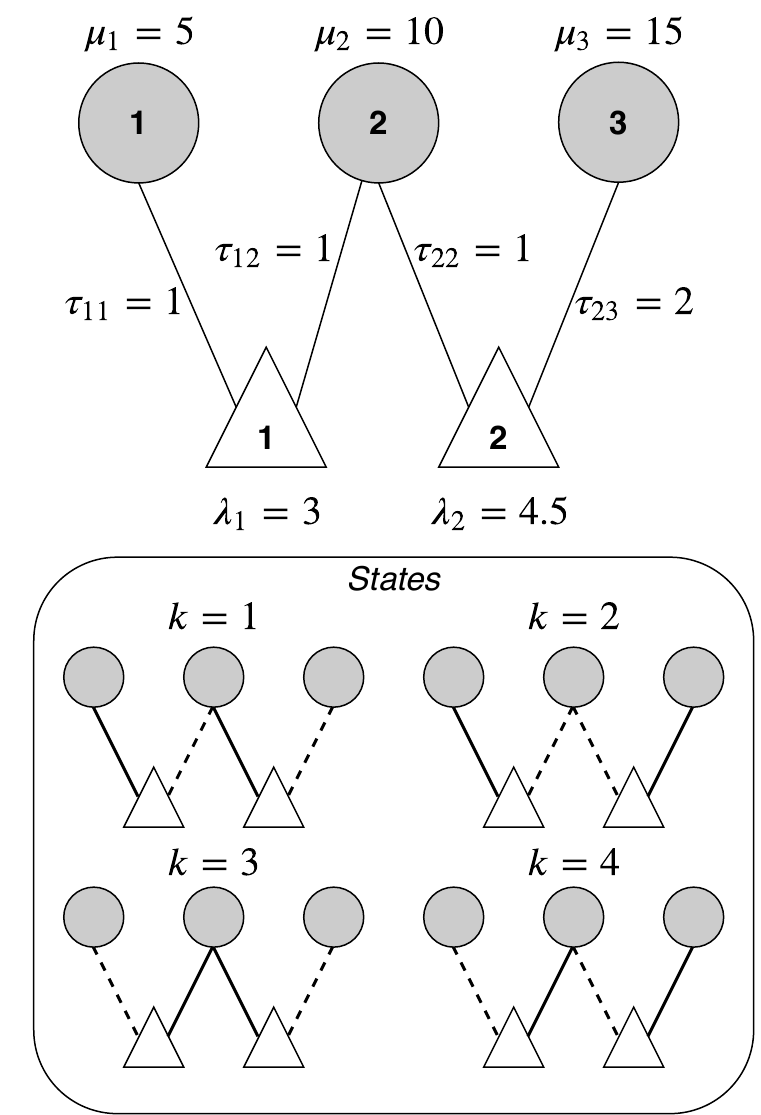}{%
Toy example of a system with two clients and three servers illustrated
in \rsec{model:example} to visualize the data structures involved
in the model.}

We now report a small numeric example, with the only purpose of
guiding the reader towards an easier understanding of the proposed
notation and model.
We have two clients (1 and 2) and three executors (1, 2, and 3),
whose arrival and serving rates, and network delays, are shown
in \rfig{example} and given below:
\begin{equation}
  \begin{array}{rcl}
    \mathbf{\lambda} & = &
    \begin{bmatrix}
      3 & 4.5  \\
    \end{bmatrix} \\
    \mathbf{\mu} & = &
    \begin{bmatrix}
      5 & 10 & 15 \\
    \end{bmatrix} \\
    \mathbf{x} & = &
    \begin{bmatrix}
      1 & 1 \\
    \end{bmatrix} \\
    \mathbf{\tau} & = &
    \begin{bmatrix}
      1 & 1 & 2 \\
      1 & 1 & 2 \\
    \end{bmatrix} \\
  \end{array}.
\end{equation}
As can be seen: client 2 has a heavier load but it can use faster
executors; the fastest executors, i.e., 3, has the highest network
delay.

First, we enumerate all possible 4 states, which are illustrated
graphically in the bottom part of \rfig{example} and formally
determined in our notation as:
\begin{equation}
  \begin{array}{rcl}
    \mathbf{s} & = &
    \begin{bmatrix}
      1 & 1 & 2 & 2 \\
      2 & 3 & 2 & 3 \\
    \end{bmatrix} \\
    \mathbf{\bar{s}} & = &
    \begin{bmatrix}
      2 & 2 & 1 & 1 \\
      3 & 2 & 3 & 2 \\
    \end{bmatrix} \\
  \end{array}.
\end{equation}
So far, we have simply defined the structures holding the \textit{input}
of our problem.
Let us now move to the analysis, starting with determining the
average delays in the current vs.\ probing destination for all
clients in all states, using \req{avg-delay-primary} and
\req{avg-delay-secondary}, respectively:
\begin{equation}
  \begin{array}{rcl}
    \mathbf{\delta} & = &
    \begin{bmatrix}
      3.5 & 3.50 & 5.00 & 2.52 \\
      2.9 & 3.42 & 5.00 & 3.42 \\
    \end{bmatrix} \\
    \mathbf{\bar{\delta}} & = &
    \begin{bmatrix}
      2.92 & 2.08 & 2.06 & 2.06 \\
      3.03 & 2.08 & 3.03 & 2.52 \\
    \end{bmatrix} \\
  \end{array}.
\end{equation}

Based on the $\mathbf{\delta}$ and $\mathbf{\bar{\delta}}$ average
delays, we can then write down all the $\mathcal{J}_k$ sets of
states that can be reached from state $k$.
For instance, $\mathcal{J}_1 = \{3\}$ because, by looking only to
the first column of $\mathbf{\delta}$ and $\mathbf{\bar{\delta}}$,
we see that for client 1 in state 1 it is better to move to its
secondary executor because $\delta_{11} > \bar{\delta}_{11}$, whereas
for the client 2 in state 1 it is better to stick to its primary
executor since it is $\delta_{21} < \bar{\delta}_{21}$; thus, the
only possible transition is from state 1 to state 3, see also the
graphical representation of the states in \rfig{example}.
Based on the formal definition of $\mathcal{J}_k$ in \req{definition-jk}
and \req{transition-prob}, we can build the transition probability
$P$ as:
\begin{equation}
  P = 
  \begin{bmatrix}
    -1    &  0   & 1    & 0    \\
    0.33  & -1   & 0.33 & 0.33 \\
    0.33  & 0.33 & -1   & 0.33 \\
    0.33  & 0.33 & 0.33 & -1   \\
  \end{bmatrix},
\end{equation}
which is irreducible and with positive recurrent states, and
it has the following stationary distribution:
\begin{equation}
  \mathbf{\pi} = 
  \begin{bmatrix}
    0.25 & 0.19 & 0.37 & 0.19 \\
  \end{bmatrix}.
\end{equation}
Eventually, the average delay per client are determined by
multiplying $\mathbf{\delta}$ by $\mathbf{\pi}$ which gives:
\begin{equation}
  E[\delta] =
  \begin{bmatrix}
    3.88 & 3.89 \\
  \end{bmatrix},
\end{equation}
therefore in this example the two clients have a very similar average
delay.

\myssec{Analysis}{model:analysis}

In this section we report a numeric analysis obtained with the model
described in \rsec{model:system} above.
The tools and scripts used are available as \textit{open source}
software on GitHub\internetref{https://github.com/ccicconetti/markovsim}.

In a first batch of results, we compare our uncoordinated serverless
access scheme to the baseline solution of statically allocating
clients to executors.
In both cases, the association between the client and its only
executor (or its primary and secondary executors) is random.
We measure the performance in terms of the average delay of clients,
which is given by \req{avg-delay-simple} for the single executor
case and by \req{delay-final} with dual executors.
We evaluate the performance as the load grows, by increasing the
number of clients from 2 to 14; on the other hand, we consider 4,
6, and 8 executors, respectively, while keeping the overall serving
rate equal to 96 (i.e., with 4 executors each has a $96/4=24$ serving
rate, etc.).
For simplicity, we consider that the network delay is negligible
for every client-executor pair.
The value of $\chi$ is always 0.1.
For each combination of parameters we ran 100 independent runs.
In each run we draw randomly the load of every client from 1 to 3,
each with the same request duration equal to 1.

\myfigeps{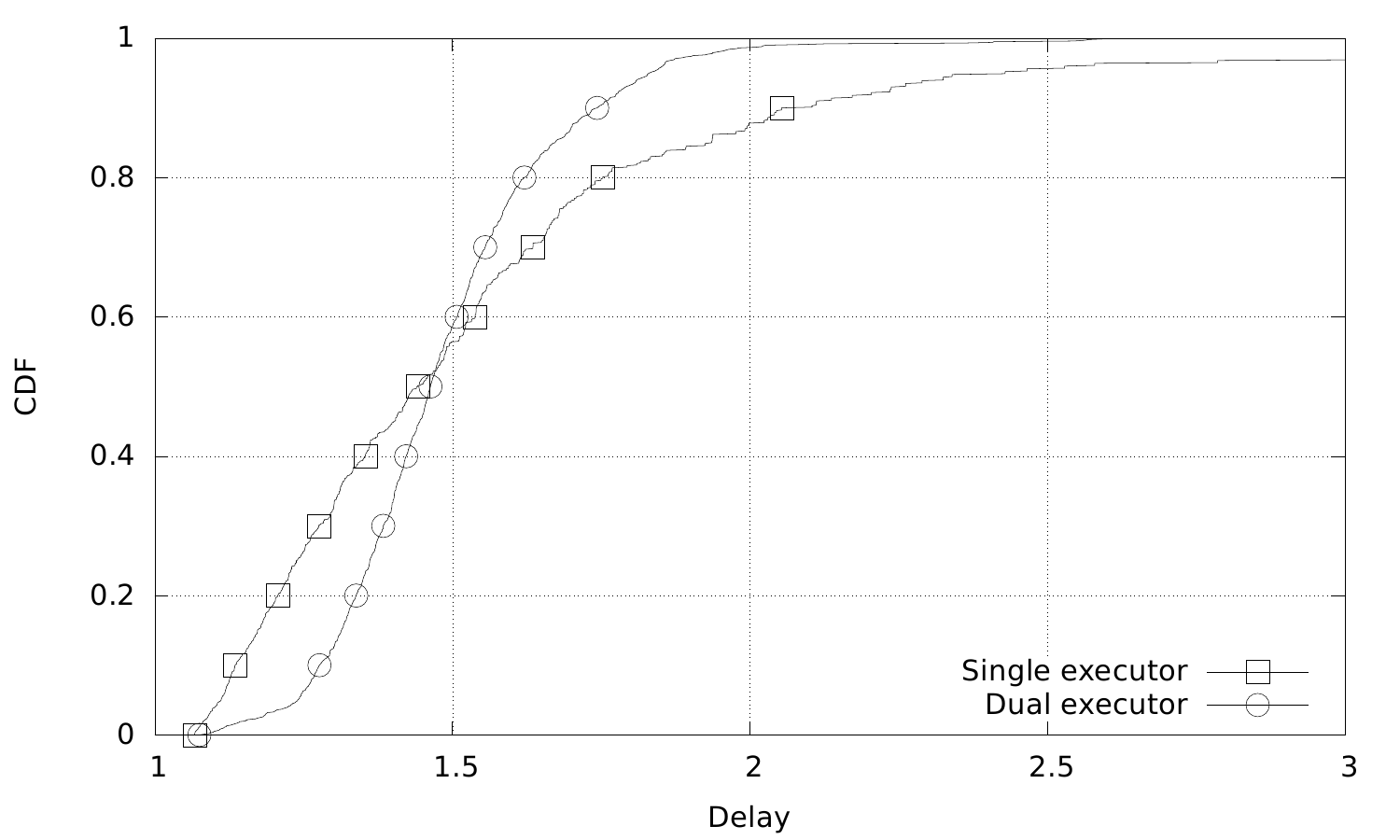}{%
Delay distribution in single vs.\ dual
executor scenarios, with 10 clients and 6 executors.}

In \rfig{single-vs-dual-dist} we plot the \ac{CDF} of the delay,
in a random but representative combination of 10 clients and 6
executors.
We can see that while the median in the two cases is almost the
same, the dual executors distribution is much less skewed than that
with a single executor: the probing mechanism in the uncoordinated
serverless access proposed is very effective in keeping the delay
of clients within a smaller range, even with a random assignment
of clients to executors, which is clearly a worst case.
This property is especially important for those \ac{IoT} applications
that rely on the response time for the execution of a remote function
being upper bounded for correct/smooth operation.

\myfigfulleps{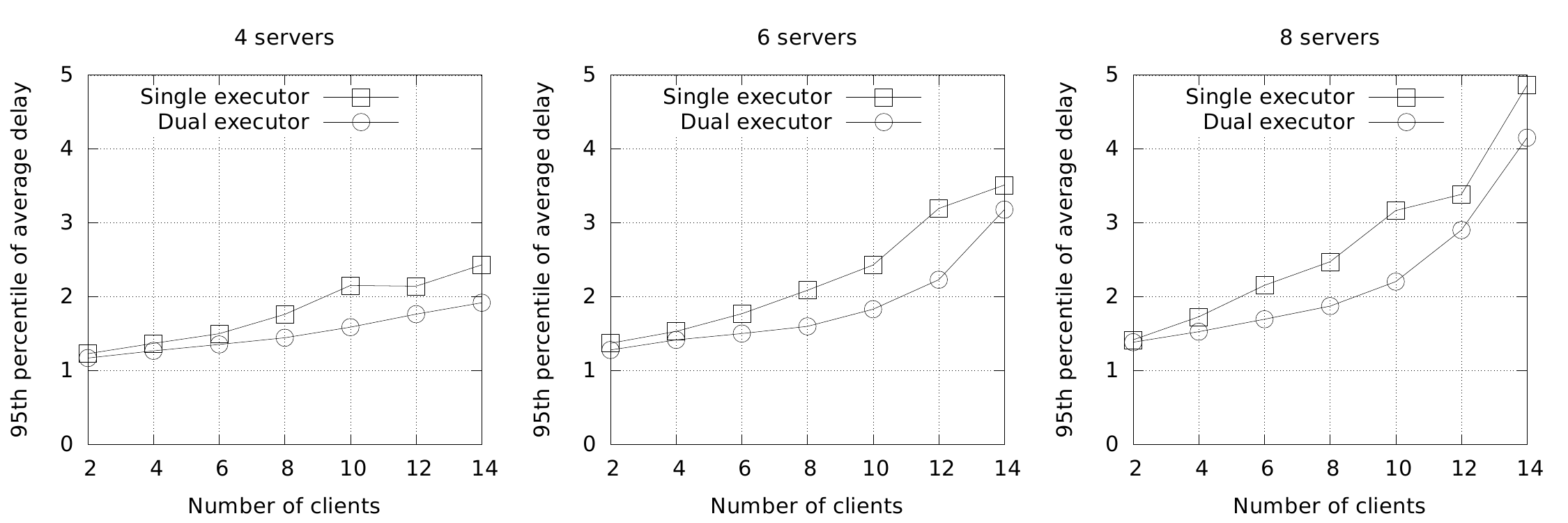}{%
Comparison between single vs.\ dual executor
with 4, 6, and 8 executors and an increasing
number of clients, in terms of the 95th
percentile of the average delay of the users.}

In \rfig{single-vs-dual-q95} we summarize the results obtained is
all the combinations studied, by reporting only the 95th percentile
of the average delay of the clients.
First of all we note that the curves decreases as the number of
executors decreases, for both the single executor and the dual
executors: since the \textit{overall } serving rate is the same,
it is expected that having a smaller number of executors reduces
the probability that a single executor becomes overloaded as a
result of an uneven allocation of clients to executors.
Second, as can be seen, the single executor curve always lies on
top of that with dual executors, which confirms the behavior described
for the specific case in \rfig{single-vs-dual-dist} in all the
combinations tested.

We now study another scenario, with the goal of assessing the
impact of $\chi$ on the system dynamics.
We keep the number of servers constant and equal to 6.
Also, the client load is equal to 2, whereas the serving rate
of the executors is drawn uniformly between 8 and 16.
We increase the number of clients from 6 to 10, and the value
of $\chi$ from 0.001 to 0.5.
In this case we ran 1000 independent replications for every combination
of the factors.

\myfigeps{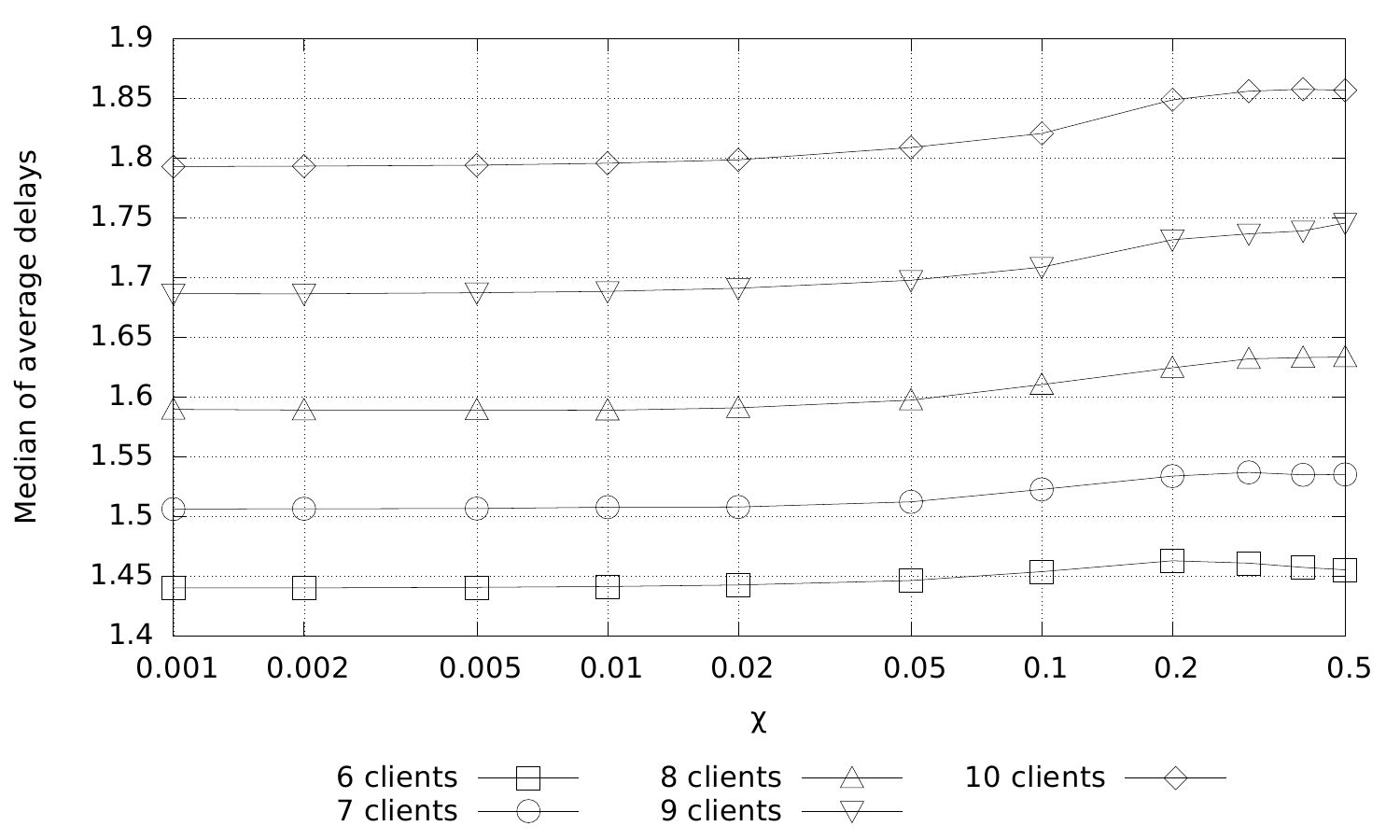}{%
Median of average delays with increasing $\chi$ with different
number of clients.}

First, in \rfig{var-chi-q050}, we show the median of the average
delays of all clients with increasing $\chi$ and number of clients.
As can be seen, the delay is not very sensitive to large changes
of $\chi$ in the range under test, which is positive because we can
expect this parameter not to have a crucial relevance in the overall
system configuration.
However, especially at higher loads, we can see that high values
of $\chi$ tend to exhibit a higher median average delay.

\myfigeps{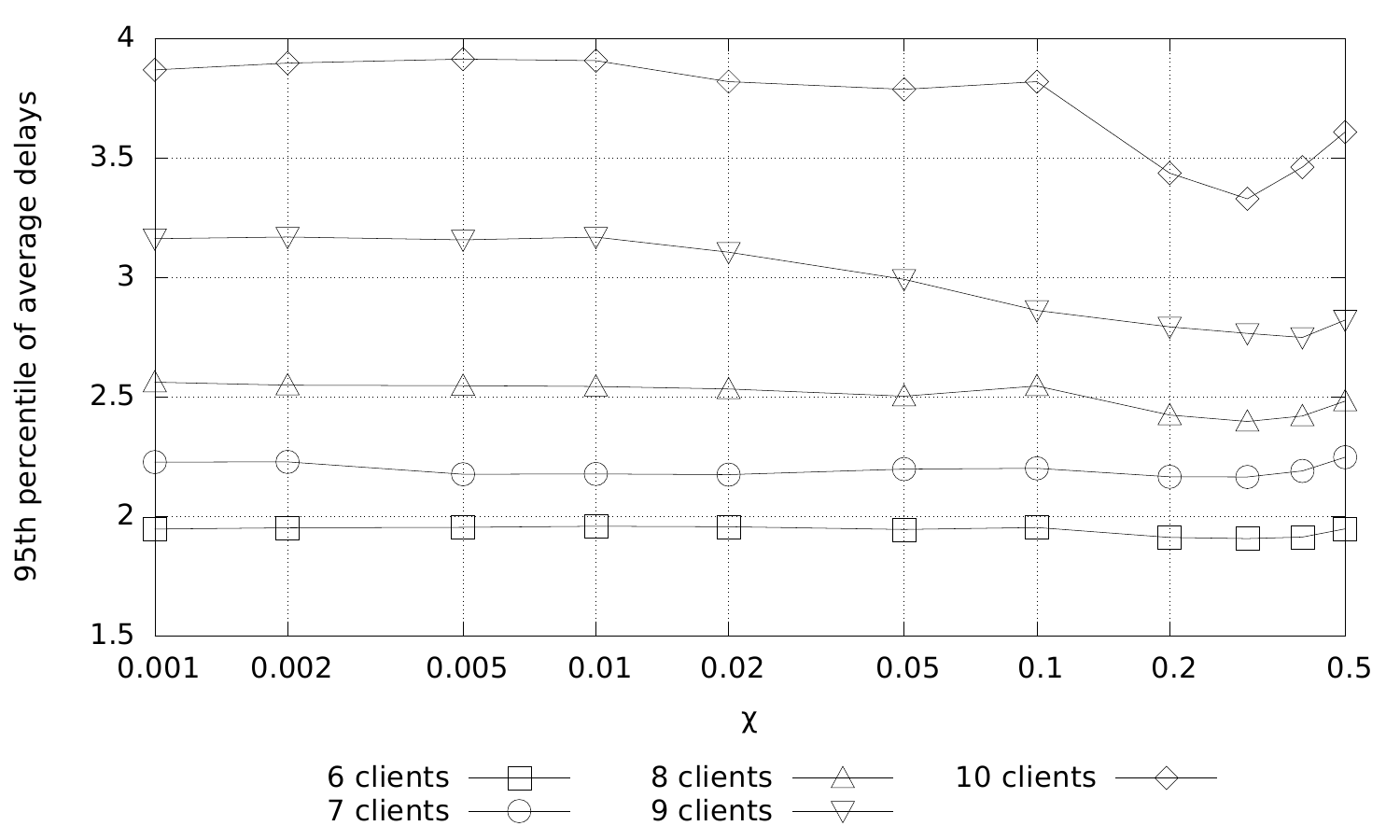}{%
95th percentile of average delays with increasing $\chi$ with
different number of clients.}

We then show the 95th percentile of the average delays in
\rfig{var-chi-q095}.
Like the median, the 95th percentile is not affected significantly
by changes of $\chi$ below 0.01.
However, with higher values, and again especially at higher loads,
the 95th percentile of the average delays decreases as $\chi$
increases.
Intuitively, the reason for this is that at high loads \textit{exploration}
becomes more important because there is a high chance that, due to
uneven allocation, one of the executors is heavily loaded.
In other words, when increasing $\chi$ the overall system load also
increases, because the clients do more probing, but, depending on
the conditions, the extra load can be useful as it benefits users
that would otherwise spend too much of their time with an overloaded
primary executor.
As can be expected, there is a trade-off: as the value of $\chi$
becomes too high, then the delay increases again because overall
the system becomes overloaded.

\myfigeps{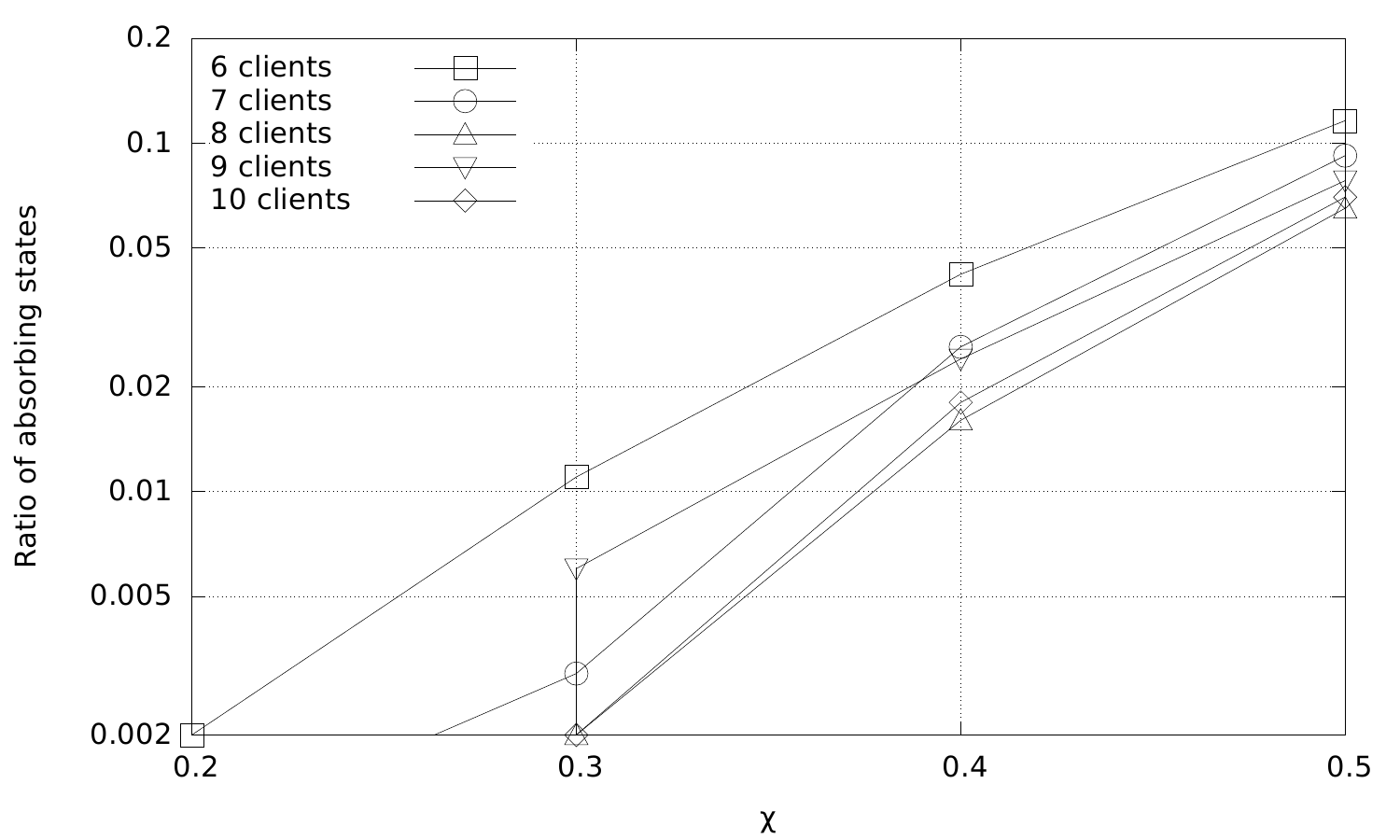}{Ratio of experiments with absorbing states.}
Another effect of the value of $\chi$ being too high is that
many states of the system have unstable queues, thus leading
to a much sparser transition matrix $P$ in our model.
This, in turn, fosters the appearance of absorbing states, which
are otherwise extremely rare, as can be seen in \rfig{var-chi-absorbing},
which shows the ratio of scenarios leading to a transition matrix
with an absorbing state over the total number of replications in
the same conditions.
With $\chi < 0.2$ the ratio is $0$, hence it is not shown in the
graph, but it increases steeply (note the y-axis logarithmic scale
in this plot) after $\chi = 0.2$, for all the number of clients.
We leave as future work deeper elaborations on predicting the
conditions leading to a transition matrix with an absorbing state
and on its system performance impact.
  \section{Performance evaluation}%
  \label{sec:eval}%
  In this section we study the performance of the proposed solution
for uncoordinated serverless access with a testbed implementation
in an emulated edge network.
We first introduce the methodology and tools used for the evaluation
(\rsec{eval:methodology}).
\removed{%
Then we discuss the results obtained in two scenarios: a small-scale
one (Sec.~5.2) to validate the qualitative conclusions
from the model analysis in Sec.~4; a large-scale
scenario in more realistic conditions (Sec.~5.3).
}%
\added{%
Then we discuss the results obtained in two scenarios aimed at
different objectives.
In \rsec{eval:small} we validate the qualitative conclusions from
the model analysis in \rsec{model:analysis} in a non-realistic
scenario that mimics the system model defined therein.
In \rsec{eval:large} we set up an environment in realistic conditions
to assess the performance of our uncoordinated serverless access
scheme, compared to alternative state-of-the-art solutions.
}

\myssec{Methodology and tools}{eval:methodology}

In this paper we re-use the performance evaluation framework
described in~\cite{Cicconetti2018}, briefly summarized in
the following.
Performance evaluation of edge systems is a challenging task:
full-scale deployments are most accurate but they require a huge
effort for the realization and may seldom be configured in such a
way to run fully repeatable experiments; cloud simulators are very
versatile but they focus on modeling adequately only one or few
aspects of the system (e.g., data placement in~\cite{Naas2018} or
scheduling in~\cite{FERNANDEZCERERO2018160}); finally, packet-level
simulators (e.g.,~\cite{Qayyum2018}) include realistic models for
the communication but cannot easily accommodate real applications.
We believe that our approach achieves a good trade-off between
accuracy of results under realistic condition and execution in a
controlled and repeatable environment, by using real applications
running in lightweight containers emulating a real network
with mininet\internetref{http://mininet.org/}.
The clients and servers are written in C++ and they communicate via
REST interfaces, realized with the popular
gRPC\internetref{https://grpc.io/} library from Google.

For scalability reasons, lambda executors do not perform computations
based on the input, but instead simply emulate the behavior of an
application running in a \ac{VM} with given virtual resources
assigned, in terms of number and speed of CPUs and amount of memory
available, processing incoming requests with a pool of pre-allocated
workers, where waiting tasks are served with a \ac{FCFS} policy.
In both scenarios below we have configured the lambda executor
emulators so that a single worker fully using its CPU requires
50~ms processing time for a lambda request of size 5,000~bytes.
We have carried out a sensitivity analysis to verify that the
conclusions are not affected by this particular choice, as well as
by some others listed in the respective scenarios (including the
lambda request rate and the number of executors).
The results are however not reported in this work because they do
not provide the reader with significant insights on the matters
under study.

We have implemented the following solutions for comparison reasons:
\begin{itemize}[leftmargin=*]
  \item \textbf{uncoordinated-2}/\textbf{uncoordinated-3}: the
  uncoordinated serverless access proposed in \rsec{contribution},
  with two and three possible destinations, respectively;
  \added{%
  based on preliminary results (included in the supplementary
  material) additional destinations yield inferior performance
  in the scenarios we have considered;
  }%
  recall that in our mathematical model in \rsec{model} we limited
  ourselves to just two destinations to keep it tractable;
  \item \textbf{static}: the allocation of every client to just one
  executor, as the baseline approach in edge computing also implicitly
  assumed by the \ac{ETSI} \ac{MEC};
  \item \textbf{centralized}: a single node in the network performs
  load balancing, using a weighted round-robin policy, where the
  weight is equal to a running estimate of the execution latency
  towards the given destination; this approach is illustrated
  at a high level in the example in the top part of \rfig{architectures};
  \item \textbf{probing}: same as centralized, but lambda
  dispatching happens by first querying all the executors on
  the processing time required if no other task is received,
  then selecting the one reporting the shortest duration; this
  approach is proposed in~\cite{Tan2017} as a solution for
  scheduling tasks in an edge-cloud system;
  \item \textbf{distributed}: same as centralized, but there are
  dispatchers distributed over the network that perform load balancing
  purely based on their local information to limit the communication
  overhead, as we proposed in~\cite{Cicconetti2018}.
\end{itemize}

Every experiment has been repeated with the same configuration, but
different seeds for the initialization of random number generators,
until achieving statistical convergence.
In the plots we report 95\% confidence intervals where appropriate
for the type of experiment and unless they are negligible compared
to the respective mean values.
Each experiment lasts for 70 seconds or 310 seconds, respectively
in \rsec{eval:small} and \rsec{eval:large}, but the initial
10\% is always considered as warm-up and the measurements in that
period are discarded.
The value of $\chi$ in all the experiments reported below is constant
and equal to $0.1$ \added{as a compromise between reacting fast to
changing conditions (which would require $\chi$ as big as possible)
and keeping the probing overhead reasonable (overhead increases
with $\chi$).
The value was found based on a preliminary analysis whose results
are not reported in the paper but available as part of the supplementary
material.}

\myssec{Small-scale scenario}{eval:small}

\myfigeps[width=2.5in]{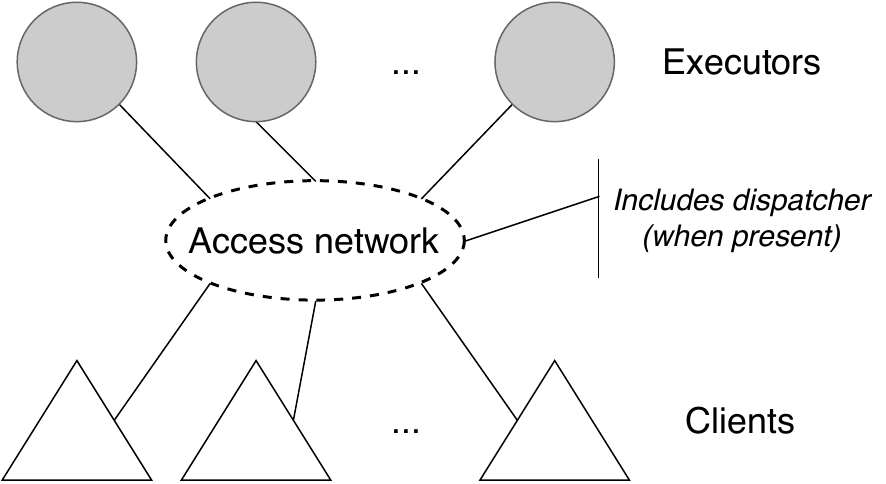}{%
Small-scale scenario: network topology.}

In this section we aim at validating the conclusions inferred from
the numerical analysis of mathematical model in \rsec{model:analysis}:
uncoordinated serverless access brings advantages, in terms of the
high percentiles of delays, compared to static allocation of clients
to executors, despite it increases the overall system load.
\added{%
Specifically, we set to achieve this goal in a topology that clearly
benefits a centralized or distributed solution (defined in
\rsec{contribution:arch}): as illustrated in \rfig{uncoord-initial-topo},
a single access network separates the clients from the edge nodes,
also providing a perfect ``natural'' location for a load balancer.
As a matter of fact, since both the centralized and distributed
policies here would have a single load balancer, there is no distinction
between them and, thus, they are identified as a single case in
plots.
}%
\removed{%
Thus we have arranged a scenario in ``artificial'' conditions, so as
to be as close as possible in our real-time testbed to the system
model assumptions.
The network topology is shown in Fig.~12 and
it is clearly a degenerate case of edge system where a single access
network separates the clients from all the edge nodes with no further
structure.
}
All links have a 100~Mb/s capacity with 1~$\mu$s delay.
The clients continuously issue an average of 5 lambda requests per
second following a Poisson distribution.
The number of servers is always 8 while the number of clients is
increased from 16 to 32, which also increases proportionally the
overall load.
In every experiment we select randomly the number of CPU cores
available per executor.
For the static, uncoordinated-2, and uncoordinated-3 policies, the
set of target destinations per executor is selected randomly in
every experiment; for the others, the load
balancing node is located in the access network, which is the
more natural placement providing best results.
\removed{%
Note that in this scenario there is no distinction between the
centralized and distributed policies, which are, thus, identified
as a single case in plots.
}

\begin{figure*}[t]%
\centering%
\includegraphics[width=0.33\textwidth]{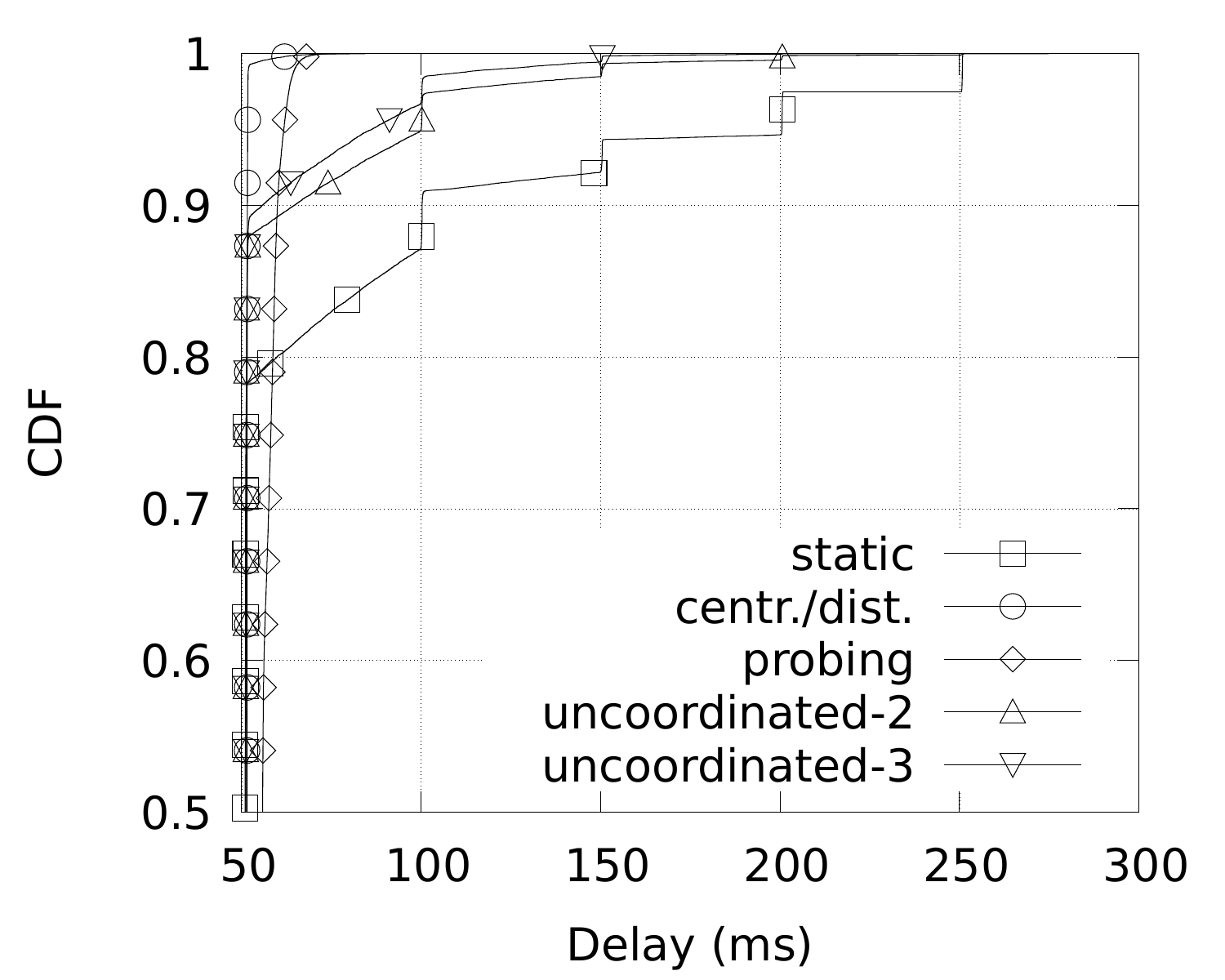}%
\includegraphics[width=0.33\textwidth]{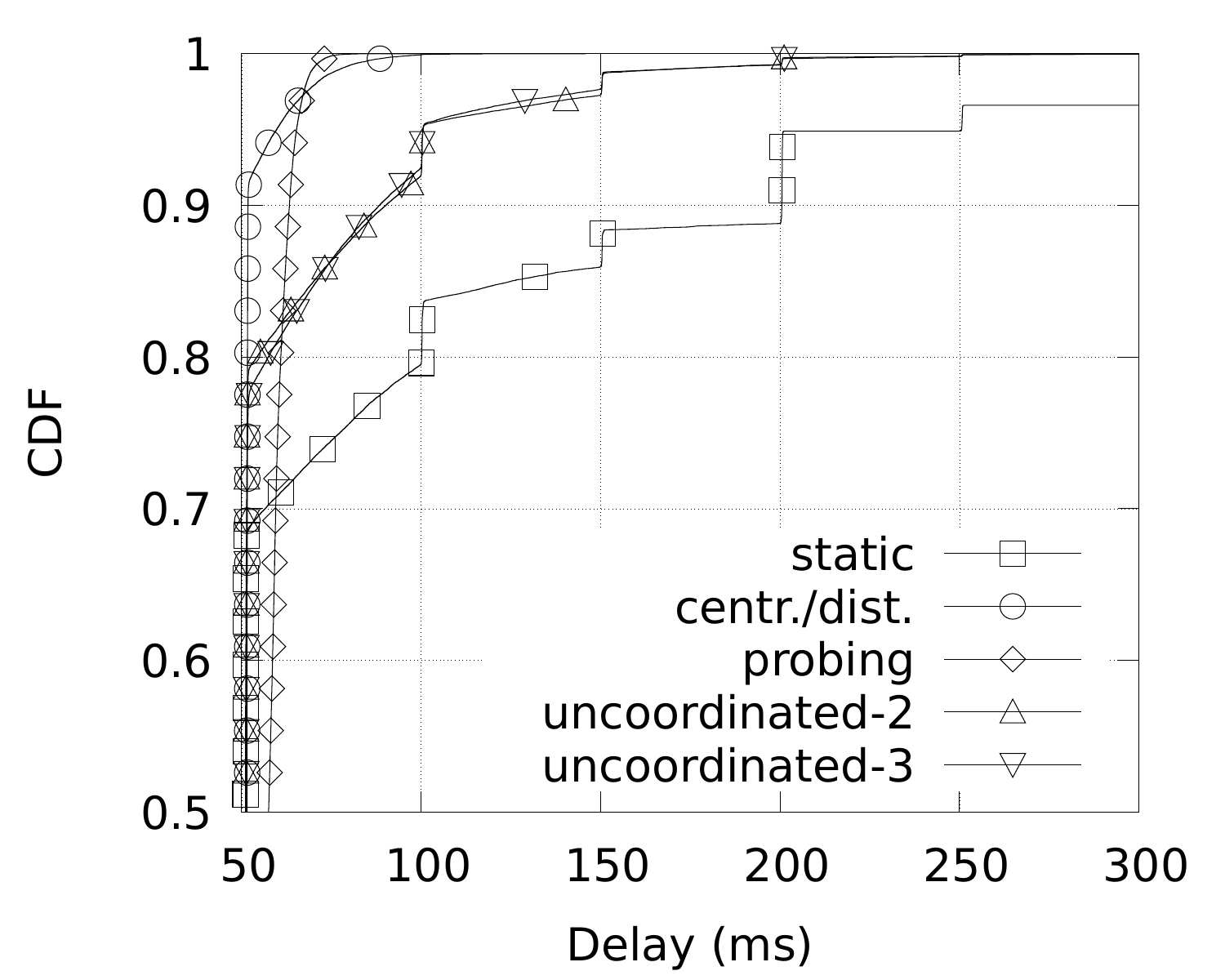}%
\includegraphics[width=0.33\textwidth]{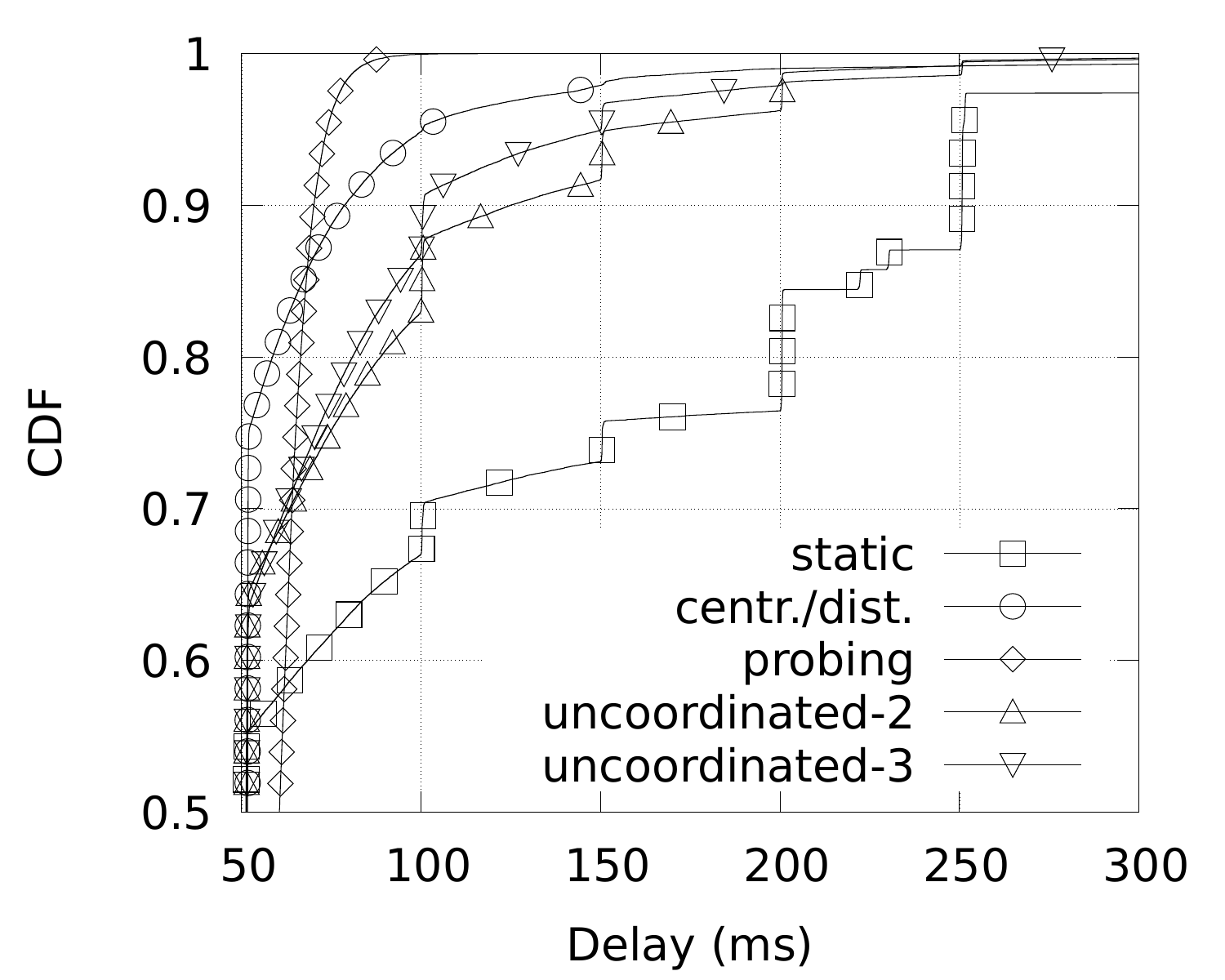}%
\caption{Small-scale scenario: delay distribution with 16 (left),
24 (center), and 32 (right) clients.}%
\label{fig:uncoord-initial-dist-poisson}%
\end{figure*}%

In \rfig{uncoord-initial-dist-poisson} we show the cumulative
distribution of the \textit{delay}, which is defined as the time between
when a client issues a lambda request towards the destination (or
the load balancer, when present) and when it fully receives back
the response.
With uncoordinated-2 and uncoordinated-3 the multiple lambda requests
are fired in parallel and the delay stops as the first response is
received from the executor requiring the least processing + networking
time, with further responses being simply discarded.
Every curve has been obtained by putting together all the
delays of all clients in all independent repetitions run
with a given combination of number of clients and policy used.
Thus, confidence intervals are not applicable to this metric.

As we can see from the plots, the probing policy achieves excellent
results at all traffic loads.
This is because the centralized entity, which in this topology
is perfectly located, asks the executors about the processing time
of every incoming lambda requests, which makes the mechanism
robust to both uneven allocation of computation resources and
temporary congestion due to unbalanced traffic.
As a matter of fact, in~\cite{Tan2017} the authors prove that such
a scheduler is $(1+\epsilon)$-speed $\mathcal{O}(1/\epsilon)$-competitive,
which is extremely good in a system where it has been proved that
no online algorithm can be optimal.
Unfortunately, this solution cannot be implemented in practice
because, in general, an executor does not know beforehand the time
required for the execution of a function.
Also, the traffic overhead caused by this approach is significant,
as will be seen later.
Thus, we consider probing merely as an ideal performance reference.

Load balancing, indicated in the plots as centr./dist., is instead
a viable solution, which yields a smooth, but relatively small,
increase of the delay as the load increases from 16 to 32 clients.
The attentive reader may have noticed that centr./dist.\ achieves
even \textit{better} performance than probing with only 16 clients:
this is because the former is not encumbered by having to ask the
executors about the future processing time, as the latter is required
to do.
The performance of our proposed uncoordinated scheme is only
marginally worse than that of centr./dist., which in our opinion
is very remarkable because it does not require any additional
architectural element that would add complexity (hence development
and maintenance costs), hamper the scalability, and become a single
point of failure, as elaborated extensively in \rsec{contribution}.
In this scenario the difference between uncoordinated-2 and
uncoordinated-3 is only slight with 32 clients and negligible at
lower loads.
Finally, a static allocation exhibits poorest performance by far:
as already evident from the results of the numerical analysis in
\rsec{model:analysis} in simplified conditions, adding just one
more destination option greatly improves the performance in terms
of delay, especially at high percentiles, which are most important
in latency-sensitive \ac{IoT} applications.

\myfigeps{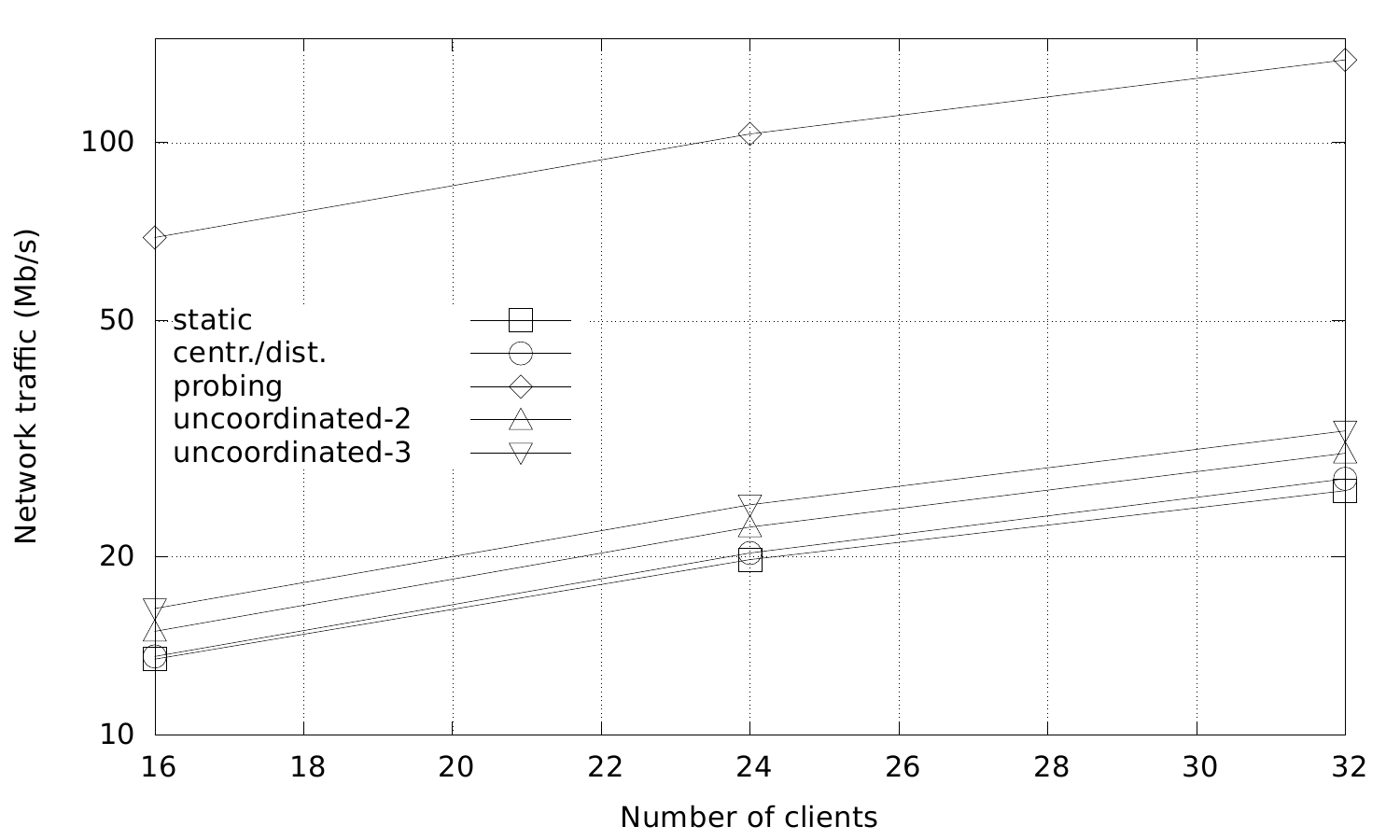}{%
Small-scale scenario: network traffic.}

In \rfig{uncoord-initial-tpt-mean} we show the overall \textit{network
traffic}, defined as the sum of the average traffic in the unit
of time of all the network links.
\added{%
In this work, the metric is an indirect measure of the overhead
incurred by the various strategies adopted: since there are no other
ongoing transmissions between nodes in our experiments, under the
same rate of lambda requests served, if solution~A has a higher
network traffic than solution~B it means that A required additional
data exchanges compared to B.
}
As introduced earlier, we see that probing has a huge network
overhead, even in such a small-scale topology as that
considered.
On the other hand, static has the lowest traffic requirement, which
is rather obvious since the clients transmit to a single executor
(unlike uncoordinated-2 and -3) and without the need to maintain
an overlay, as with a distributed approach.
The uncoordinated solutions exhibit a slightly higher network
traffic, which creates the following trade-off: the higher the
number of destinations (and the higher the value of $\chi$), the
lower are high quantiles of delay but the higher is the communication
overhead.
Depending on the target environment and expected \ac{QoS} requirements
of the applications, a suitable calibration must be done by the
edge system operator to achieve best performance.

\myssec{Large-scale scenario}{eval:large}

\myfigeps{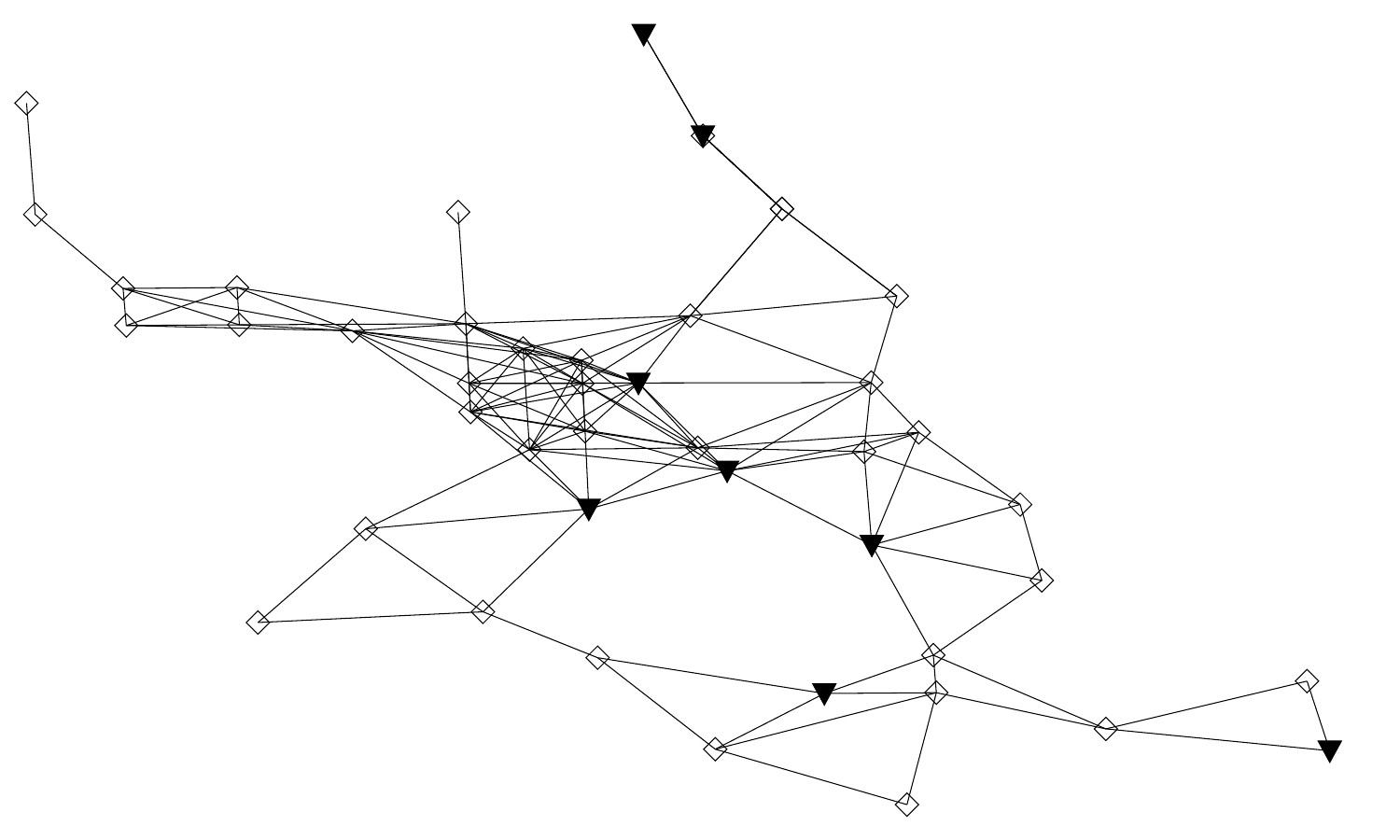}{%
Large-scale scenario: sample network topology showing
servers (circles) and clients (triangles), both
also acting as intermediate networking devices.}

In this section we use a large scale scenario in a topology extracted
from a real \ac{IoT} network: Array of Things%
\internetref{https://arrayofthings.github.io/}, a collaborative
effort with about 100~nodes installed at intersections in Chicago,
IL, US, using the Waggle platform~\cite{7808975}, which is more
realistic than both the system model in \rsec{model:system} and the
environment in the previous experiments (\rsec{eval:small}).
Starting from the geographical locations of the real nodes, we
have first collapsed nodes that are too close to one another, then
added bi-directional 100~Mb/s capacity / 1~$\mu$s latency links
between nodes based on a threshold distance.
The resulting network map is illustrated in~\rfig{uncoord-chicago-topo}
and it consists of 45~nodes with a diameter of 11~hops.

In this scenario the clients adopt the following traffic pattern:
a burst of lambda requests is generated at the beginning of consecutive
periods, whose duration is drawn from an exponential distribution
with mean 10~s.
The burst size, expressed in terms of number of lambda requests,
is drawn from a Poisson distribution with mean 25.
After receiving the response, the client backs off for a random
amount of time, drawn from a uniform distribution in $[150, 200]$~ms
before issuing the next request, to model processing on the client
side.
Like in the previous scenario, the executor emulators and client
applications are configured in such a way that a single task
requires 50~ms to be executed on a given core with no other
concurrent task being processed.
Thus, the duty cycle at low loads is about 0.5 (= $25 \times (50 +
150) / 10^3$).

In every replication, 8 out of the 45 nodes are selected as
executors, each with two CPU cores and workers.
All the nodes may host clients, which are dropped randomly
at the beginning of every experiment.
The number of clients is increased from 40 to 80.
In uncoordinated-2 and -3, the destinations are selected among those
having a shortest path from the client; for instance, with
uncoordinated-2 if there is destination A three hops away and
destinations B, C, and D four hops away, then the rest are farther,
we select randomly two destinations out of \{A, B, C, D\}.
By extension, with static we always select the closest executor,
breaking ties randomly when required.
With centralized we select randomly the node acting as
load balancer and all the clients contact the executors
through the latter.
With distributed we assume that all the nodes hosting an executor
also host a distributed dispatcher; clients always contact the
closest dispatcher for the execution of lambda requests.
\added{%
In this scenario, when using a probing policy the system becomes
unstable, i.e., the traffic consumed by the central load balancer
for polling all the executors to determine which one is best
suited to serve the next incoming lambda request is so high
that the communication links are saturated, which leads to
ever-growing queues (and delays) of client requests.
This confirms the impossibility to implement the probing policy
in a practical scenario.
Results with probing are not shown in plots.
}
\removed{%
In this scenario it is not meaningful to use a probing policy,
since the network overhead nullifies every advantage coming
from the non-realistic querying done towards the executors.
}

\myfigeps{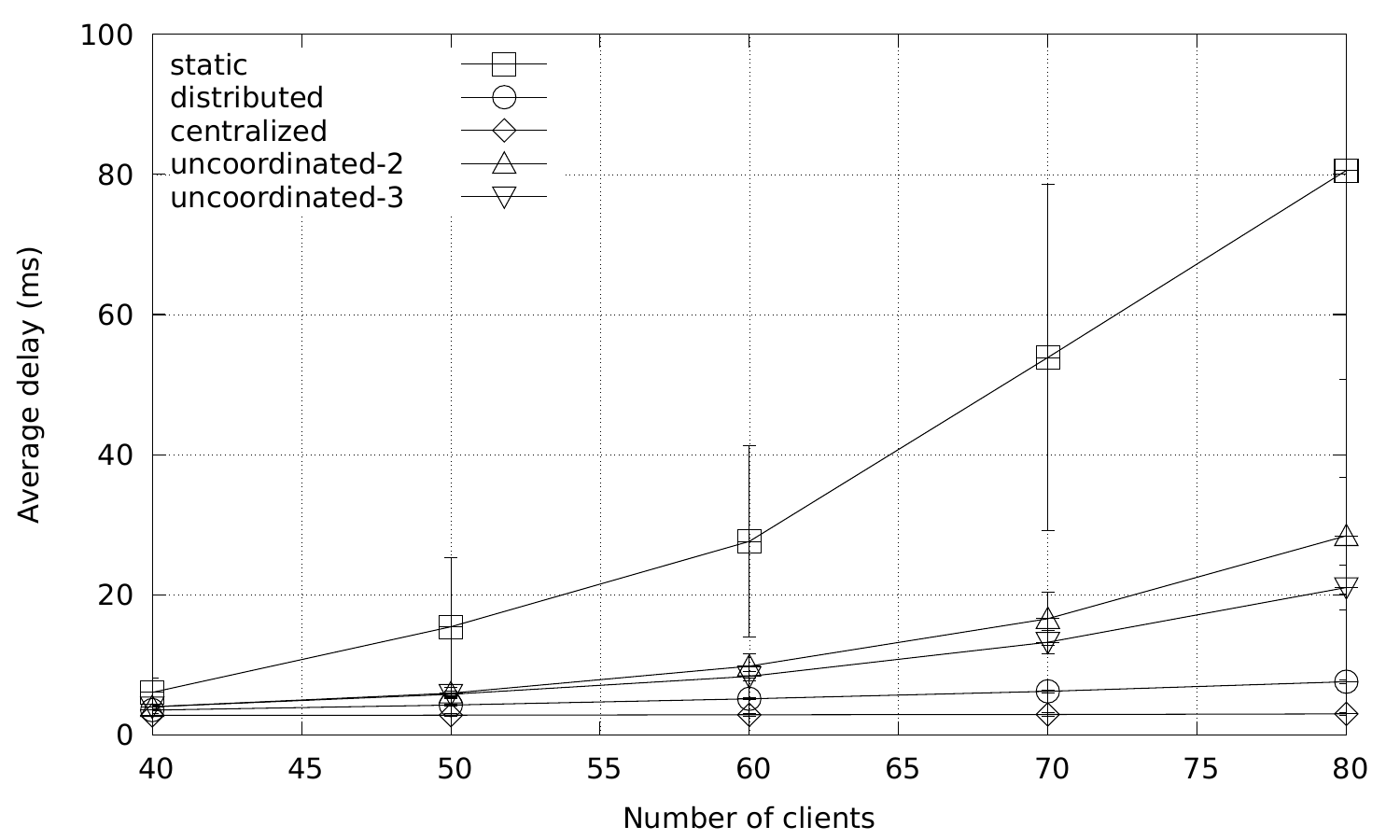}{%
Large-scale scenario: average delay.}

\myfigeps{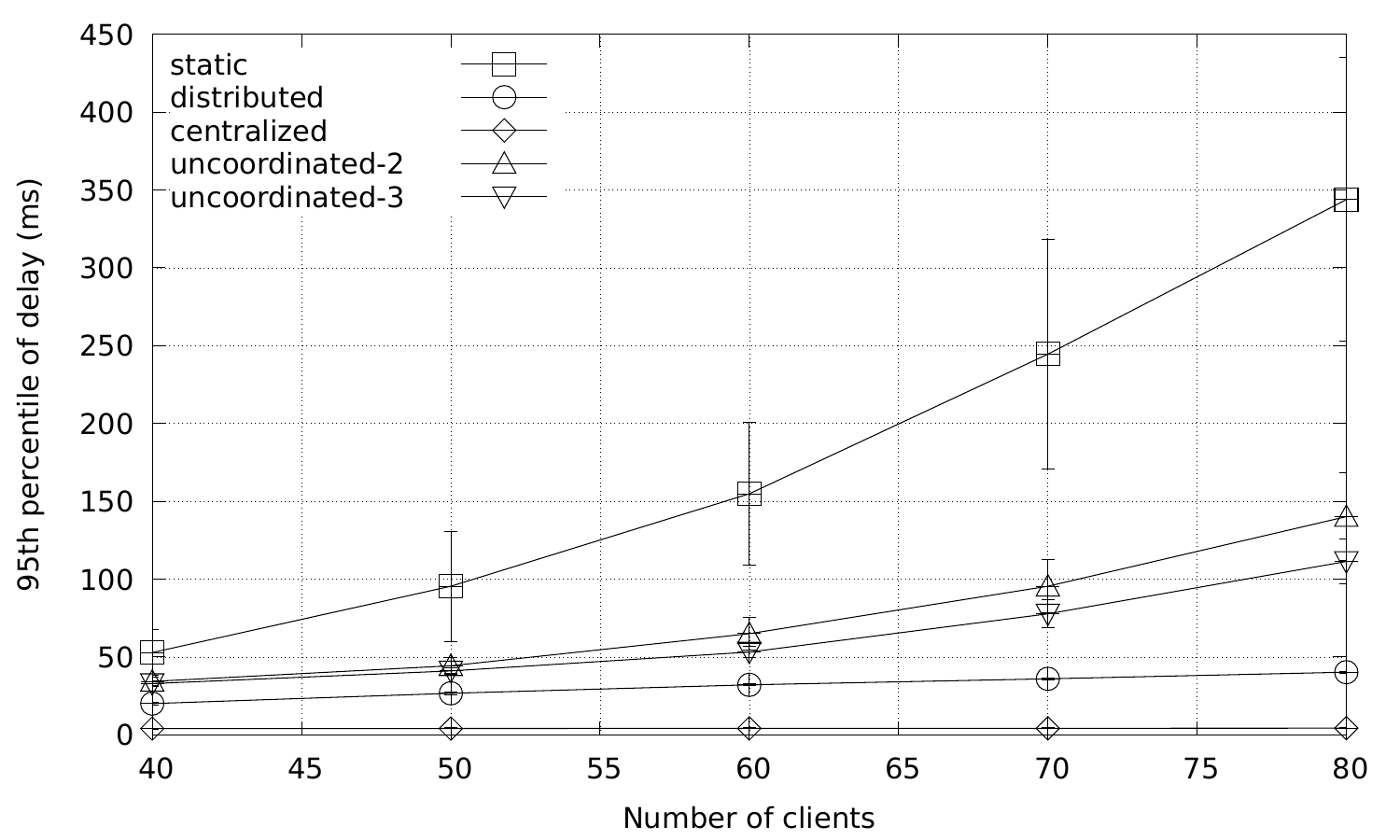}{%
Large-scale scenario: 95th percentile of delay.}

\added{%
As in the previous section, the key performance index for this
scenario is the delay, which in this section is subtracted a constant
value equal to the minimum processing time of the lambda requests,
i.e.,\ 50~ms, for better readability of plots.
%
}
\removed{
We begin with analyzing the jitter, which is defined as
the difference between the delay (as in Sec.~5.2 above)
and the minimum processing time, i.e. 50~ms.
}
\removed{
The goal is to keep the jitter as low as possible to have a consistent
response time for all the applications of the same type in the
network, irrespective of their location and the current state of
the system.
}
As can be seen in \rfig{uncoord-chicago-out-mean}, uncoordinated-2
and -3 achieve intermediate performance in terms of the average
delay, rather close to that of distributed and centralized, while
the static curve lies well above the rest.
This behavior is exacerbated for the 95th percentile of the
delay, reported in \rfig{uncoord-chicago-out-095}.
This confirms that also in a more realistic topology with
bursty traffic an uncoordinated serverless access provides
a significant advantage, in terms of delay, compared
to a static allocation of clients to executors.
Performance can be improved further by using more sophisticated
policies, which however require new components and have a
higher network overhead.

\myfigeps{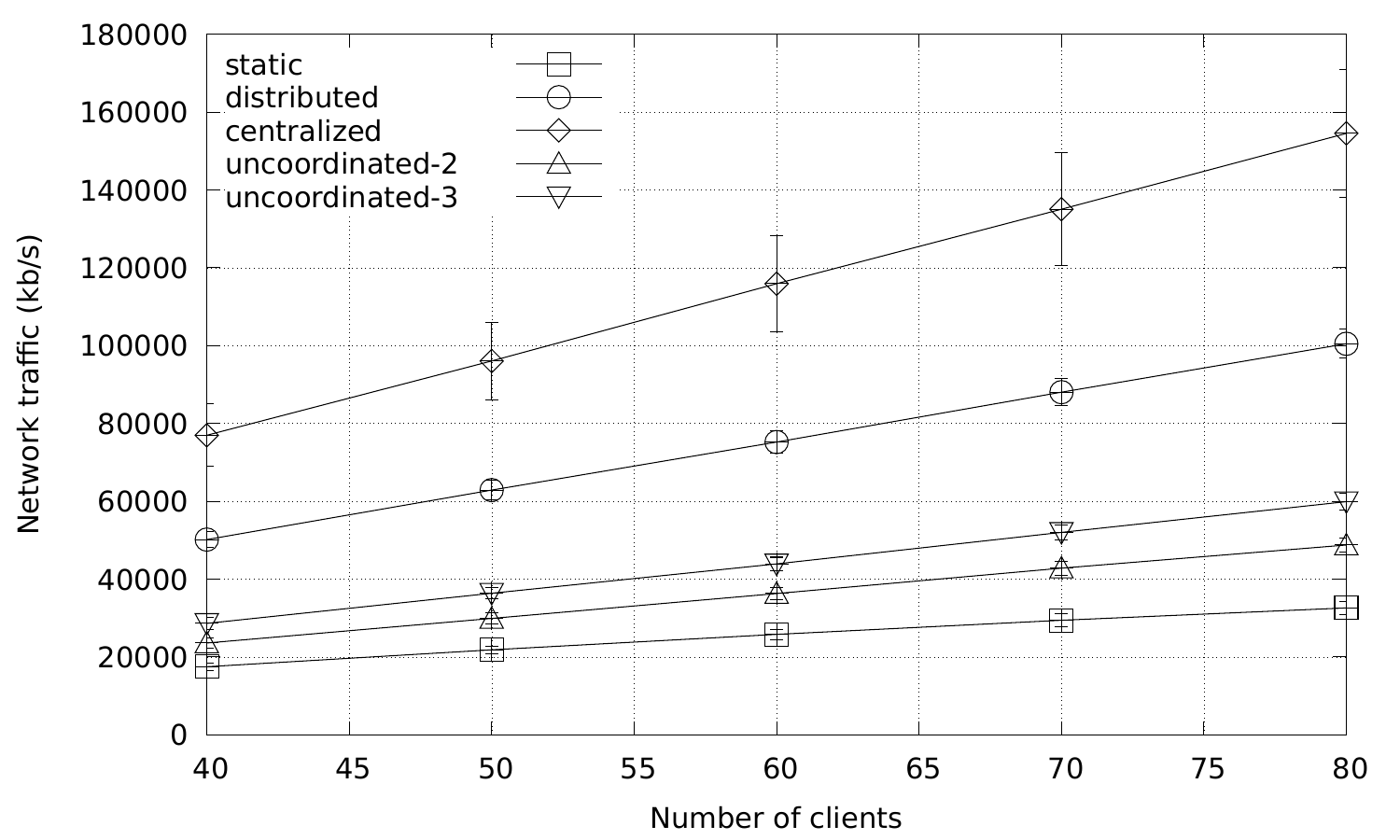}{%
Large-scale scenario: network traffic}

The last statement is proved in \rfig{uncoord-chicago-tpt-mean},
which shows the overall network traffic.
Unlike the previous scenario, which was very optimistic for
the centralized/distributed policies, the network overhead of
both is significantly higher than that of uncoordinated access.
This is because, without a natural central node in the network,
the use of an overlay for dispatching lambda tasks can be very
expensive.
Instead, the price to be paid by uncoordinated-2 and -3 compared
to static, in terms of network traffic, is limited, and deemed to
be affordable in most cases because of the advantages it brings in
terms of delay and reliability.

\myfigeps{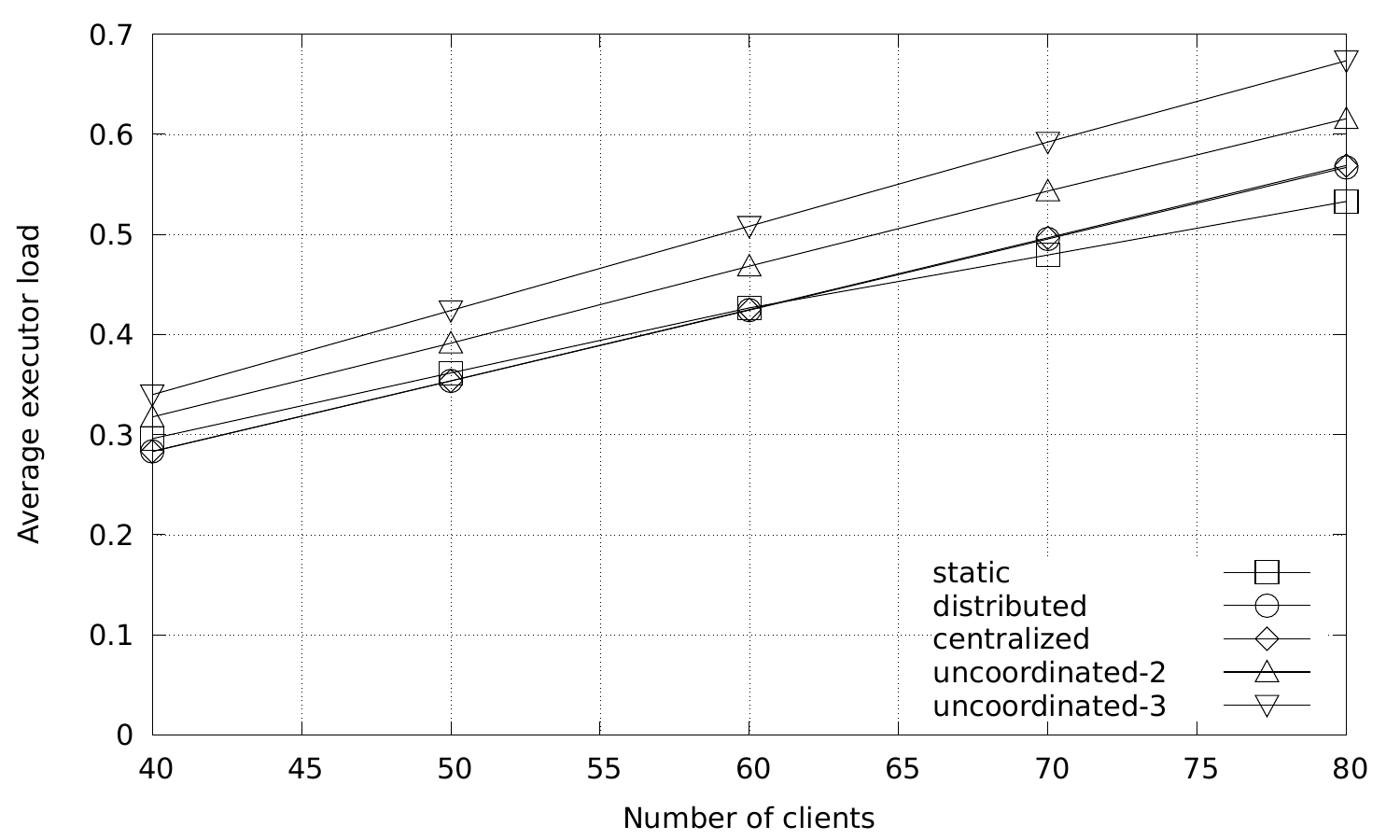}{%
Large-scale scenario: average executor load.}

We conclude the analysis with the average \textit{utilization} of the
executors, defined as the ratio between the time an executor is
busy processing at least one task and the experiment duration, in
\rfig{uncoord-chicago-util-mean} (confidence intervals here are
omitted because negligible).
This is to show that uncoordinated access requires an additional
price: computational resources on the executors must be invested
to process lambda requests without a strict necessity to do so.
In fact, while distributed, centralized and static have almost 
overlapping performance, the utilization becomes higher with
uncoordinated-2 and even more so with uncoordinated-3.
We leave for future studies the design of more sophisticated policies
that retain most advantages of the uncoordinated access techniques,
while also reducing their network and computational overhead.
  \section{Conclusions and future work}%
  \label{sec:conclusions}%
  In this paper we have investigated the problem of fast changing
connectivity and computational load conditions for serverless
\ac{IoT} applications running in an edge network.
Based on a critical analysis of state-of-the-art solutions relying on
either centralized dispatching or a distributed overlay, we have
proposed a new approach called \textit{uncoordinated serverless
access}, which does not require complex/costly/fragile system elements,
hence it is practical to implement in fast growing and fragmented
\ac{IoT} deployments.
We have developed a mathematical model using queuing theory under
simplified assumptions, which shows that the proposed approach
reduces the \removed{jitter}\added{delay} of response times compared
to a static allocation of clients to micro-service executors, which
is today's baseline.
These numerical results have been confirmed by experiments carried
out with a prototype implementation in two emulated networks, one
of which uses a realistic topology and bursty traffic.
The goal achieved is especially important for latency-sensitive
applications, which can be found in many areas of huge practical
interest, such as connected car and industry automation.
In the emulation experiments we have also compared uncoordinated
serverless access with centralized load balancing and distributed
dispatching: the results have shown that our proposed solution, in
addition to being simpler and requiring fewer maintenance, requires
much less network traffic ($ -65\%$ than centralized, $-55\%$ than
distributed) while requiring only $+10\%$ computational load on
executors.
Finally, the uncoordinated access scheme proposed can be realized
within the \ac{ETSI} \ac{MEC}.\removed{ with minor changes to the \texttt{Mx2}
interface, which have been described in the paper.}

Even though distributing computing has been extensively studied in
the scientific literature and is a mainstream technology for cloud
systems, very few of the models and technologies apply to \ac{IoT}
systems, especially when used in edge networks, which are emerging
as the most viable approach to a sustainable deployment in several
business areas.
In our opinion, what we have presented in this paper is only the
beginning in a new area of research on how to design, operate,
optimize, and maintain complex systems where \ac{IoT} devices
consume services offered by heterogeneous devices close to them
with limited compute and connectivity capabilities.

With specific reference to this work, we believe it could be extended
in at least the following directions: further elaboration on the
mathematical model to infer actionable properties in some specific
conditions (e.g., with homogeneous population of clients); closed-loop
systems to modify at run-time the system parameter configuration
(e.g., $\chi$ or the number of destinations per client); more
sophisticated stateful algorithms (e.g., including prediction/estimation)
to be used by clients that are powerful enough; integration with
orchestration systems for an optimized selection of the destinations
of each client beyond shortest-path.
%


\begin{acronym}
  \acro{ML}{Machine Learning}
  \acro{3GPP}{Third Generation Partnership Project}
  \acro{5G-PPP}{5G Public Private Partnership}
  \acro{AA}{Authentication and Authorization}
  \acro{AP}{Access Point}
  \acro{API}{Application Programming Interface}
  \acro{AR}{Augmented Reality}
  \acro{ARP}{Address Resolution Protocol}
  \acro{BGP}{Border Gateway Protocol}
  \acro{BS}{Base Station}
  \acro{CDF}{Cumulative Distribution Function}
  \acro{CPU}{Central Processing Unit}
  \acro{DTMC}{Discrete-Time Markov Chain}
  \acro{EPC}{Evolved Packet Core}
  \acro{ETSI}{European Telecommunications Standards Institute}
  \acro{FCFS}{First Come First Serve}
  \acro{FaaS}{Function as a Service}
  \acro{FSM}{Finite State Machine}
  \acro{GPU}{Graphics Processing Unit}
  \acro{HPC}{High Performance Computing}
  \acro{HTTP}{Hyper-Text Transfer Protocol}
  \acro{ICN}{Information-Centric Networking}
  \acro{IoT}{Internet of Things}
  \acro{IIoT}{Industrial Internet of Things}
  \acro{ILP}{Integer Linear Programming}
  \acro{IP}{Internet Protocol}
  \acro{IPP}{Interrupted Poisson Process}
  \acro{ITS}{Intelligent Transportation System}
  \acro{ITU}{International Telecommunication Union}
  \acro{KPI}{Key Performance Indicator}
  \acro{LCM}{Life Cycle Management}
  \acro{LHS}{left-hand-side}
  \acro{LL}{Link Layer}
  \acro{LTE}{Long Term Evolution}
  \acro{MBWA}{Mobile Broadband Wireless Access}
  \acro{MCC}{Mobile Cloud Computing}
  \acro{MAC}{Medium Access Layer}
  \acro{MEC}{Multi-access Edge Computing}
  \acro{MEO}{MEC Orchestrator}
  \acro{MEPM}{MEC Platform Manager}
  \acro{ML}{Machine Learning}
  \acro{MNO}{Mobile Network Operator}
  \acro{NAT}{Network Address Translation}
  \acro{NFV}{Network Function Virtualization}
  \acro{OF}{OpenFlow}
  \acro{OS}{Operating System}
  \acro{OSPF}{Open Shortest Path First}
  \acro{OWC}{OpenWhisk Controller}
  \acro{PMF}{Probability Mass Function}
  \acro{PS}{Processor Sharing}
  \acro{PU}{Processing Unit}
  \acro{QoE}{Quality of Experience}
  \acro{QoS}{Quality of Service}
  \acro{RHS}{right-hand-side}
  \acro{RPC}{Remote Procedure Call}
  \acro{RPC}{Remote Procedure Call}
  \acro{RR}{Round Robin}
  \acro{RSU}{Road Side Unit}
  \acro{TCP}{Transmission Control Protocol}
  \acro{TSN}{Time-Sensitive Networking}
  \acro{SDN}{Software Defined Networking}
  \acro{SoC}{System on Chip}
  \acro{SMP}{Symmetric Multiprocessing}
  \acro{SRPT}{Shortest Remaining Processing Time}
  \acro{UDP}{User Datagram Protocol}
  \acro{URL}{Uniform Resource Locator}
  \acro{UT}{User Terminal}
  \acro{VANET}{Vehicular Ad-hoc Network}
  \acro{VIM}{Virtualization Infrastructure Manager}
  \acro{VM}{Virtual Machine}
  \acro{VNF}{Virtual Network Function}
  \acro{VPU}{Vision Processing Unit}
  \acro{VR}{Virtual Reality}
  \acro{WLAN}{Wireless Local Area Network}
  \acro{WRR}{Weighted Round Robin}
  \acro{WSN}{Wireless Sensor Network}
\end{acronym}

\end{document}


\newpage
\renewcommand\thefigure{S-\arabic{figure}}
\renewcommand\thetable{S-\arabic{table}}
\setcounter{page}{1}
\setcounter{figure}{0}
\setcounter{table}{0}

\newcommand{\myfig}[1]{%
\begin{center}
  \includegraphics[width=0.75\textwidth]{figures/#1}
\end{center}
}

\thispagestyle{empty}

\vspace*{\fill}
\begin{center}
{\large \textbf{Supplementary material}}
\end{center}
\vspace*{\fill}

{\small%
\noindent%
\begin{tabularx}{\textwidth}{lX}
  Ref.\ code: & \texttt{COMNET\_2019\_1191} \\
  Manuscript: & Uncoordinated Access to Serverless Computing in MEC Systems for IoT \\
  Authors: & \textit{Claudio Cicconetti, Marco Conti, \& Andrea Passarella} \\
  & Institute of Informatics and Telematics --- CNR --- Italy \\
  Submitted to: & Elsevier Computer Networks --- VSI:\@ 5G-enabled IoT \\
\end{tabularx}
}

\newpage

\section*{Selection of $\chi$}

In the following we report the CDF of delay as $\chi$ increases
from 0.001 to 0.5, under the same conditions of results included
in Sec.~4.3.

\myfig{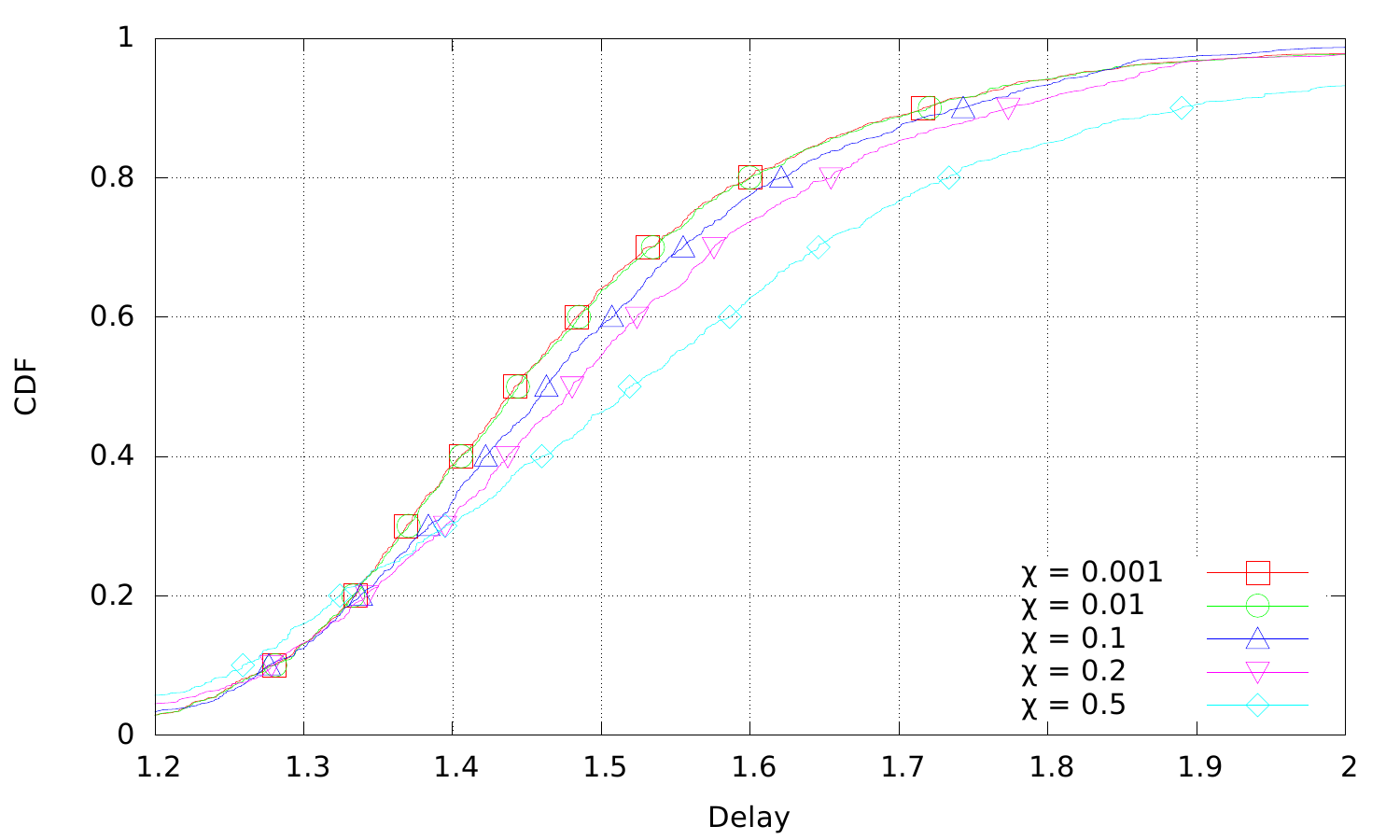}

As can be seen, the CDF of the delay varies significantly
for values greater than or equal to 0.1, while decreasing
from 0.1 to 0.01 yields a small (but still noticeable) difference,
and reducing $\chi$ below 0.01 has a negligible effect.

In practice, if we think from the point of view of a selfish device,
then $\chi$ should be as big as possible, because this means that
the device may react fast to changing conditions, since the probability
of probing a secondary executor is high.
However, since we also want to share the finite resources fairly
among all the clients, we must be parsimonious on $\chi$: too high
values lead to a penalty for all.
For this reason, in Sec.~5 we have selected the value of $\chi =
0.1$ as a reasonable trade-off between (selfish) aggressive reaction
to changing conditions and (communal) utilisation of resources.

\section*{Experiments with more than three destinations}

We now report the results obtained in the large-scale scenario
(Sec.~5.3) with four and six destinations, called uncoordinated-4
and uncoordinated-6, respectively.
These results are compared with those obtained under uncoordinated-2
and uncoordinate-3 policies, which are reported in the paper.

\myfig{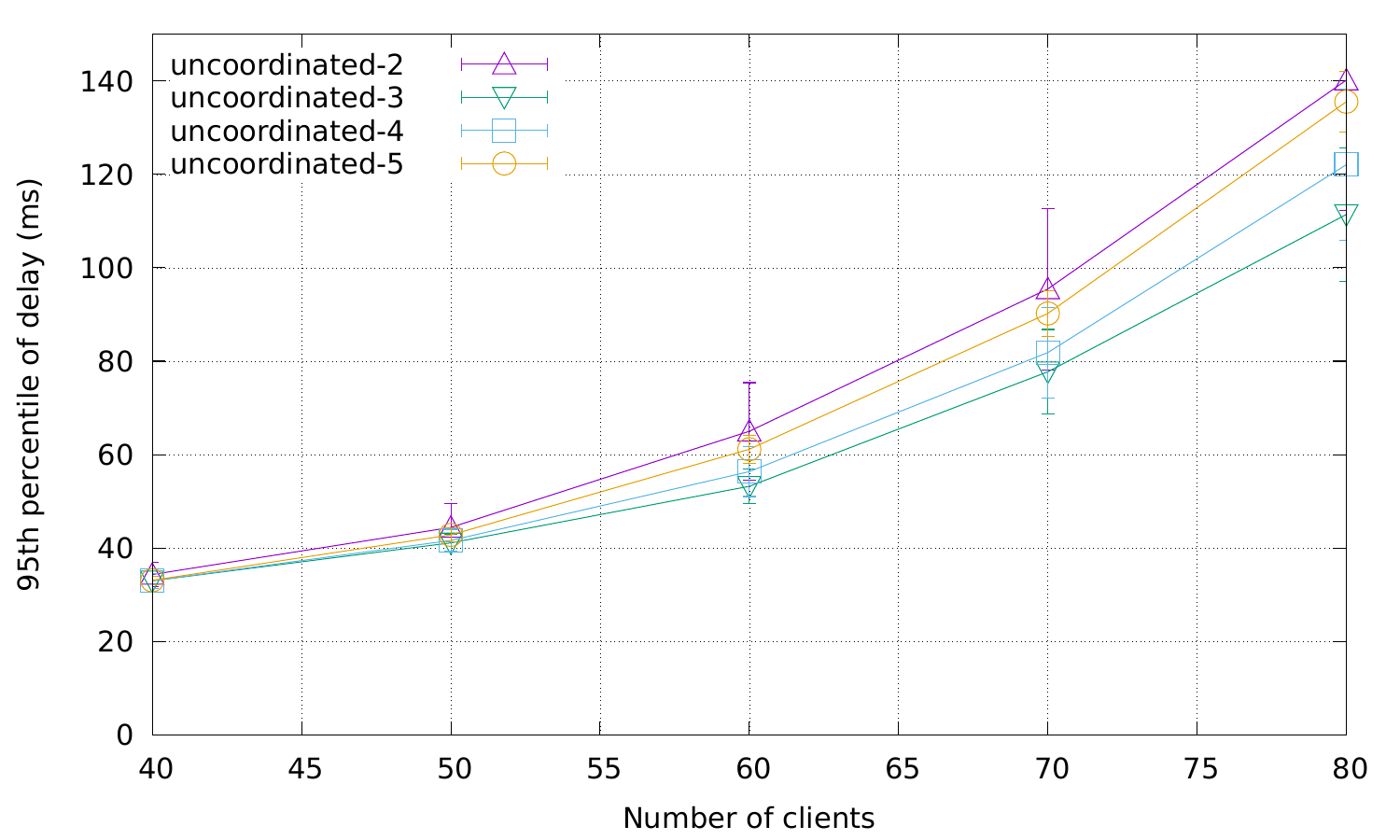}

\myfig{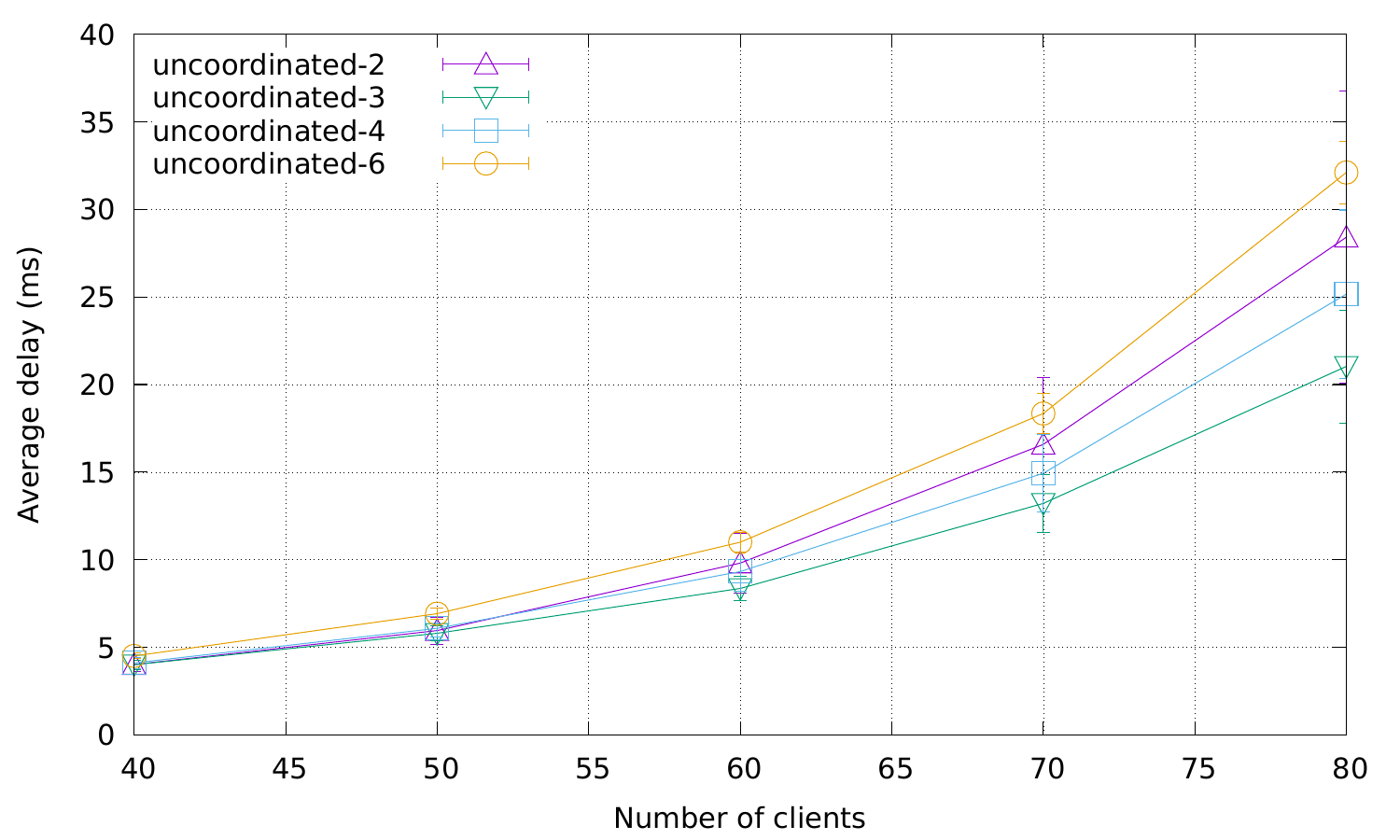}

\myfig{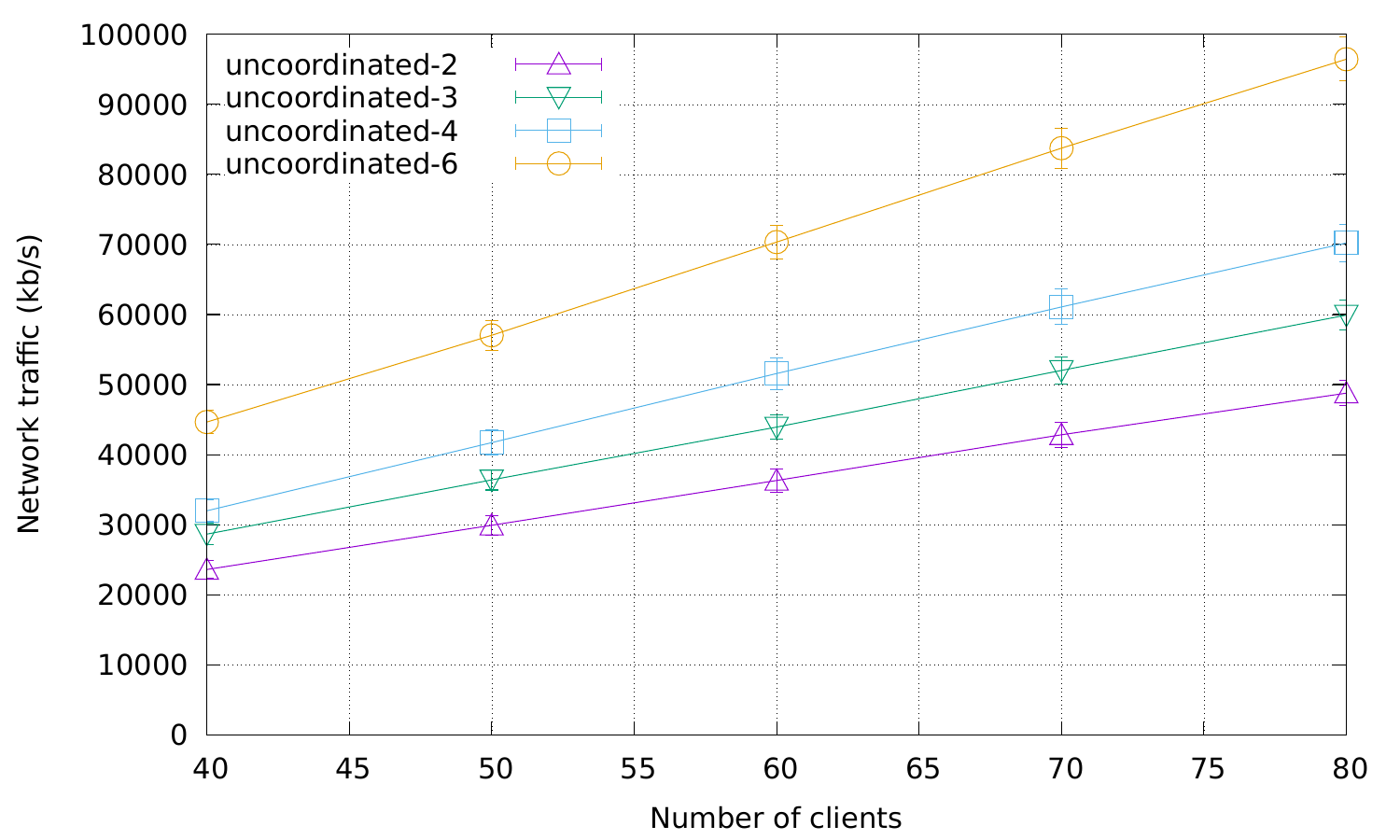}

\myfig{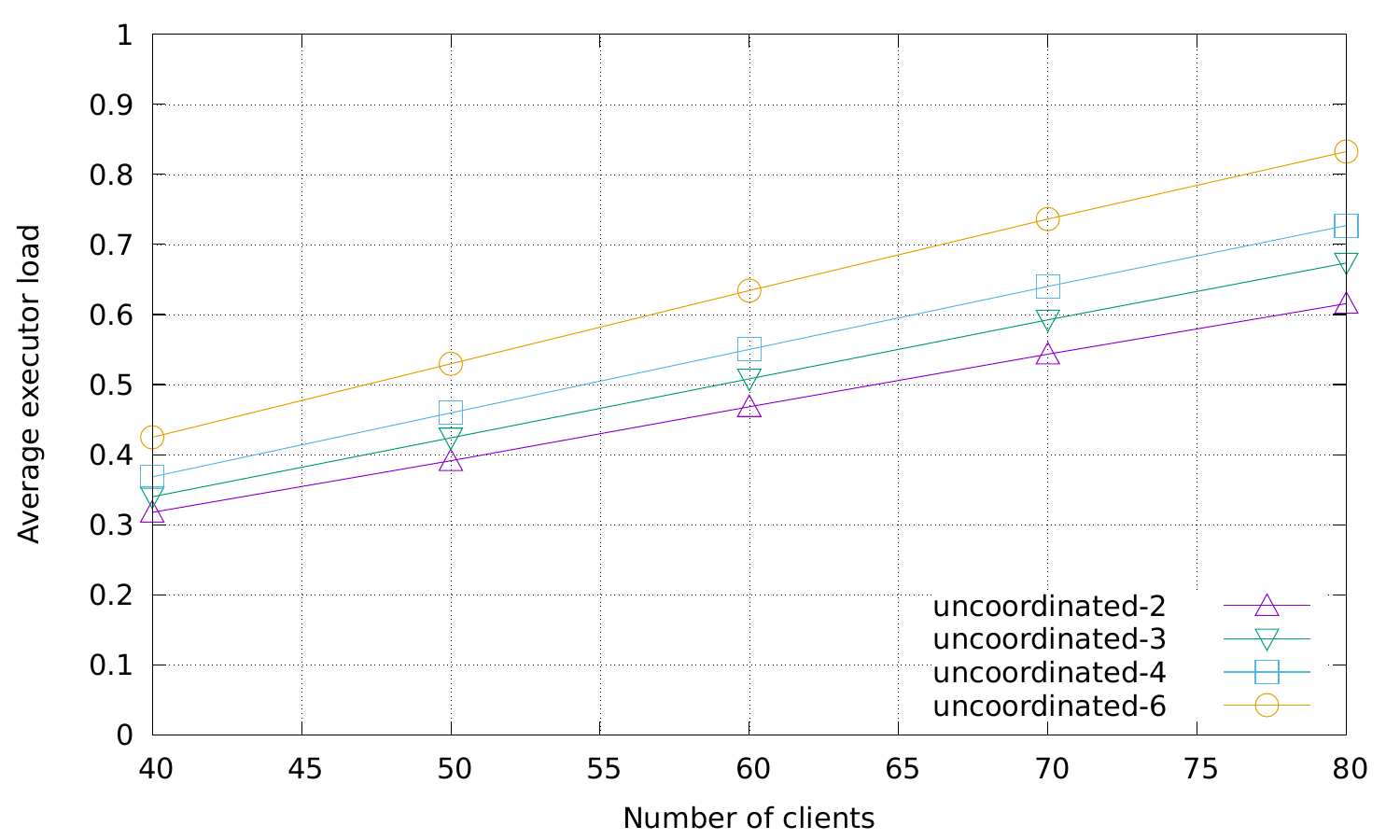}

As can be seen:

\begin{itemize}
  \item the delay, both average and 95th percentile, decreases from
  uncoordinated-2 to uncoordinated-3, because of the benefits brought
  by having a wider choice of executors in the pool; however, when
  adding further executors the delay increases again, because of the
  overhead generated at both networking and computational level;

  \item this consideration is confirmed by the network traffic and
  average executor load plots, where all the curves increase
  monotonically as the pool size increases, i.e., from uncoordinated-2
  to uncoordinated-6.
\end{itemize}

Since uncoordinated-2 and uncoordinated-3 exhibit the best trade-off
of performance, compared to uncoordinated policies with a larger
pool of executors, we have restricted the results in Sec.~5.3 of
the paper to them.